\documentclass[journal,10pt]{IEEEtran}
\usepackage{graphicx}   %
\usepackage{bm}
\usepackage{amsmath}
\usepackage{amsthm}
\usepackage{amsfonts}
\usepackage{color}
\usepackage{ulem}
\usepackage{stfloats}
\usepackage{float}
\usepackage{algpseudocode}  
\usepackage{algorithmicx,algorithm}
\usepackage{cite}
\usepackage{makecell}
\usepackage{amssymb}
\usepackage{float}
\usepackage{relsize}
\usepackage{subcaption}
\usepackage[colorlinks,linkcolor=black,citecolor=black]{hyperref}

\newcommand{\mymat}[1]{{\bm{{\rm{{}#1}}}}}
\newcommand{\defequ}{\stackrel{\triangle}{=}}

\newtheorem{remark}{Remark}

\ifCLASSINFOpdf

\else
 
\fi

\begin{document}
%
\title{\LARGE Polar-Coded Tensor-Based Unsourced Random Access with Soft Decoding}

%

\author{Jiaqi~Fang,
	    Yan~Liang,
	    Gangle~Sun,
	    Hongwei~Hou,
	    Yafei~Wang,
	    Li~You,~\IEEEmembership{Senior Member,~IEEE,}
	    and Wenjin~Wang,~\IEEEmembership{Member,~IEEE}
	    	\thanks{\textit{(Jiaqi Fang and Yan Liang contributed equally to this work.) (Corresponding
	    		author: Wenjin Wang.)}}
\thanks{The authors are with the National Mobile Communications Research Laboratory, Southeast University, Nanjing 210096, China and also with the Purple Mountain Laboratories, Nanjing 211111, China (e-mail: jiaqifang@seu.edu.cn;  yan\_liang@seu.edu.cn; sungangle@seu.edu.cn; hongweihou@seu.edu.cn; wangyf@seu.edu.cn; lyou@seu.edu.cn; wangwj@seu.edu.cn).} }     

\markboth{Journal of \LaTeX\ Class Files,~Vol.~14, No.~8, August~2015}%
{Shell \MakeLowercase{\textit{et al.}}: Bare Demo of IEEEtran.cls for IEEE Journals}

\maketitle
\begin{abstract}	
The unsourced random access (URA) has emerged as a viable scheme for supporting the massive machine-type communications (mMTC) in the sixth generation (6G) wireless networks. Notably, the tensor-based URA (TURA), with its inherent tensor structure, stands out by simultaneously enhancing performance and reducing computational complexity for the multi-user separation, especially in mMTC networks with a large numer of active devices. However, current TURA scheme lacks the soft decoder, thus precluding the incorporation of existing advanced coding techniques. In order to fully explore the potential of the TURA, this paper investigates the Polar-coded TURA (PTURA) scheme and develops the corresponding iterative Bayesian receiver with feedback (IBR-FB).
Specifically, in the IBR-FB, we propose the Grassmannian modulation-aided Bayesian tensor decomposition (GM-BTD) algorithm under the variational Bayesian learning (VBL) framework, which leverages the property of the Grassmannian modulation to facilitate the convergence of the VBL process, and has the ability to generate the required soft information without the knowledge of the number of active devices. Furthermore, based on the soft information produced by the GM-BTD, we design the soft Grassmannian demodulator in the IBR-FB. Extensive simulation results demonstrate that the proposed PTURA in conjunction with the IBR-FB surpasses the existing state-of-the-art unsourced random access scheme in terms of accuracy and computational complexity.

\end{abstract}


%
\IEEEpeerreviewmaketitle

\section{Introduction}

\IEEEPARstart{W}{ith} the approaching era of the Internet-of-Things (IoT), it is anticipated that the sixth generation (6G) wireless networks will provide massive machine-type communication (mMTC) services, such as sensing, monitoring, and reporting \cite{p_mass_access,p_gaozhen_add_1,p_gaozhen_add_2}. Generally, the mMTC possesses the characteristics of massive connectivity, sporadic traffic, and small payload, where conventional grant-based random access schemes may lead to excessive signaling overhead, network congestion, and high transmission latency \cite{p_yuwei_mmtc,p_mmtc3,p_next_noma, p_yuwei,p_lg_2}. To overcome these issues, the grant-free random access has been introduced into the mMTC to manage the massive access \cite{p_mmtc_angle,p_angle_enhance,p_lg_3}. Without the need for prior authority in the grant-based random access, the grant-free random access allows the active devices to transmit to the base station (BS) directly, thereby significantly alleviating the aforementioned problems.

Currently, there are two types of grant-free random access schemes, i.e., the sourced random access (SRA) scheme \cite{p_yuwei,p_mmtc_angle,p_angle_enhance, p_sun_g, p_zhu_y,p_mmtc_joint} and the unsourced random access (URA) scheme \cite{p_or_ura,p_low_ura,p_ccs_ura,p_sparc_ura,p_cd_ura,p_ccs_ura_angle,p_ccs_ura_nonb,p_ccs_ura_amp}. 
In the SRA scheme, every device is assigned a unique and non-orthogonal pilot for the purpose of the active user detection (AUD) and the channel estimation (CE). 
For the mMTC network with a substantial number of devices, maintaining the accuracy of the AUD and the CE necessitates the long pilot sequences, which occupy most of the packet length and lead to the suboptimal utilization of the finite system resources.
In contrast, the URA scheme is more promising, where all devices utilize a common codebook instead of being assigned unique pilot sequences. In the URA scheme, the responsibility of the BS is to restore the set of the transmitted messages, irrespective of the identities of the active devices. As the URA scheme does not necessitate the explicit initial access phase, it allows full system scalability in the sense that the system remains unaffected by the total number of devices in the network.
These advantages contribute to the improved efficiency and make the URA scheme particularly appropriate for the mMTC network involving a large number of devices.

The initial investigation on the URA is reported in \cite{p_or_ura}, where a random coding bound for the Gaussian multiple access channel (GMAC) is discussed. Following this study, a low-complexity URA scheme based on the combination of compute-and-forward
and coding for a binary adder channel is proposed in \cite{p_low_ura}.
Nevertheless, the codebook size in \cite{p_low_ura} increases exponentially with the messages length, resulting in intolerable decoding complexity.
To address this issue, many works have devoted in the low-complexity and efficient URA schemes, primarily falling into three categories: the coded compressed sensing (CCS) URA scheme \cite{p_ccs_ura,p_sparc_ura,p_cd_ura,p_ccs_ura_angle,p_ccs_ura_nonb,p_ccs_ura_amp}, the two-phase URA scheme \cite{p_polars_gmac_ura,p_polar_gmac_ura,p_idma_ura, p_ldpc_gmac_ura,p_pilot_ura,p_fas_ura}, and the tensor-based URA (TURA) scheme \cite{p_xu_ura_miso,p_xu_ura_mimo,p_tb_ura}.
In the CCS URA, messages of devices are partitioned into multiple segments for transmission, substantially diminishing the size of the codebook for each segment \cite{p_ccs_ura,p_sparc_ura,p_cd_ura, p_ccs_ura_angle}. In \cite{p_ccs_ura_nonb}, a covariance-based non-Bayesian decoder is proposed for the CCS URA. By jointly devising the inner and outer decoding process, a fully Bayesian decoder is proposed for the CCS URA in \cite{p_ccs_ura_amp}. On the other hand, messages of devices in the two-phase URA are bifurcated into two sections, where the first section is encoded by a common compressed sensing (CS) codebook and the rest is encoded
by the forward-error-correction (FEC) code with the
key parameters conveyed by the first section \cite{p_polars_gmac_ura,p_polar_gmac_ura}. The low-density parity-check (LDPC)-coded two-phase URA is developed in \cite{p_idma_ura, p_ldpc_gmac_ura}, and the Polar-coded two-phase URA is introduced in \cite{p_pilot_ura}. In \cite{p_fas_ura}, the Polar coding and the random spreading modulation are employed for the two-phase URA, leading to the fading spread URA (FASURA). 

Different from the CCS URA and the two-phase URA, the TURA is a non-segmented scheme, in which the FEC coded messages of devices are modulated and mapped as tensor symbols for transmission \cite{p_xu_ura_miso}.
Under conditions of substantial number of active devices in the mMTC network, the false alarm probability of the CCS URA escalates to intolerable levels due to the limitations imposed by the error detection capability of the parity check bits \cite{p_ccs_ura_angle, p_ccs_ura_nonb, p_ccs_ura_amp}, while the two-phase URA performance is suboptimum due to the pilot collisions \cite{p_pilot_ura, p_fas_ura}.
Comparatively, the TURA offers a more compelling solution for scenarios involving a large number of active devices, as the inherent tensor structure in it not only boosts performance but also reduces computational demands for multi-user separation  \cite{p_xu_ura_miso,p_xu_ura_mimo,p_tb_ura,p_tb_bound_ura}.
In \cite{p_xu_ura_mimo}, the sparse code is introduced into the TURA, resulting in the sparse Kronecker-product (SKP) coding scheme. By employing the Bose–Chaudhuri–Hocquenghem (BCH) code and the Grassmannian modulation for the TURA, the tensor-based modulation (TBM) scheme is proposed in \cite{p_tb_ura}. 

However, existing practical TURA is prevented from adopting current advanced coding methods due to its lack of the soft decoder, which consequently restricts its effectiveness.
To thoroughly explore the potential of the TURA, we investigate the TURA with soft decoding. Our primary contributions are summarized as follows:

\begin{enumerate}
\item 
We propose the Polar-coded TURA (PTURA) scheme. The Polar codes with exceptional error-correcting capabilities for the short messages are utilized. Moreover, the Grassmannian modulation is employed in the PTURA, which ensures the accuracy of the demodulation \cite{p_tb_ura}.


\item 
We propose the iterative Bayesian receiver with feedback (IBR-FB) for the PTURA. During each iteration of the IBR-FB, the proposed Grassmannian modulation-aided Bayesian tensor decomposition (GM-BTD) module provides soft information to the proposed soft Grassmannian demodulator, which generates the log-likelihood ratio (LLR) corresponding to each bit of the Polar codewords. Then, the messages of devices are recovered using the cyclic redundancy checking (CRC)-aided successive cancellation list (SCL) Polar decoder. Subsequently, the recovered valid messages, which satisfies the CRC requirement, are fed back to the GM-BTD module for the next iteration.
Simulation results show that the proposed PTURA in conjunction with the IBR-FB exhibits superior performance compared to the existing state-of-the-art URA scheme, especially in scenarios with a substantial number of active devices, while maintaining reduced computational complexity.

\item
We propose the GM-BTD algorithm in the IBR-FB. Specifically, the initial step in the IBR-FB is formulated as a tensor decomposition (TD) problem, which includes the canonical polyadic decomposition (CPD) as a special case. The soft decoding of the PTURA requires the TD module capable of generating soft
information, which can not be realized by the conventional
optimization-based algorithms. To address
this issue, we propose the GM-BTD algorithm under
the variational Bayesian learning framework, which can
generate the required soft information without the need
for the number of active devices. Furthermore, the GM-BTD capitalizes on the property of the Grassmannian modulation to facilitate the convergence of the variational Bayesian learning process towards the ground truth values. Simulation results show that the proposed GM-BTD algorithm outperforms the baseline algorithm.

\end{enumerate}

The rest of this paper is organized as follows:
Section \ref{sec_system_model} introduces the system model. The proposed IBR-FB is provided in section \ref{sec_iter_receiver}. In section \ref{sec_bayes_td}, we elaborate the proposed GM-BTD algorithm. Section \ref{sec_simulation} presents extensive simulation results and section \ref{sec_conclusion} concludes this paper.

\textit{Notations:}
Except for specially noted, lower and upper bold fronts, $\mymat{x}$ and $\mymat{X}$, denote vectors and matrices, respectively; $\mymat{X}[i,:]$, $\mymat{X}[:,j]$, and $\mymat{X}[i,j]$ denote the $i$-th row of $\mymat{X}$, the $j$-th column of $\mymat{X}$, and the $(i,j)$-th entry of $\mymat{X}$, respectively; $|\cdot|$, $||\cdot||_2$, and $||\cdot||_F$ denote the absolute value, the $l_2$ norm, and the Frobenius norm, respectively; $\{\cdot\}^*$, $\{\cdot\}^T$, and $\{\cdot\}^H$ denote the conjugate, the transpose, and the conjugate transpose, respectively; The statistical expectation is denoted by $\mathbb{E}\{\cdot\}$; $\Re\{\cdot\}$ and $\Im\{\cdot\}$ denote the real part and the imaginary part, respectively; $\circledast$, $\odot$, and $\circ$ denote the Hadamard product, the Khatri-Rao product, and the outer product, respectively;
$\mathbb{R}$ and $\mathbb{C}$ denote the real number field and the imaginary number field, respectively;
$\mymat{x}\sim\mathcal{CN}(\mymat{x};\mymat{\mu},\mymat{\Omega})$ denotes the complex Gaussian distributed vector $\mymat{x}$ with the mean $\mymat{\mu}$ and the covariance matrix $\mymat{\Omega}$; $x\sim\mathcal{U}(a,b)$ denotes 
$x$ is uniformly ditributed between $a$ and $b$; ${\rm{tr}}(\mymat{X})$ denotes the trace of $\mymat{X}$;
$\{\mymat{X}_l:l=1,...,L\}$ denotes a set composed of $\mymat{X}_1,\mymat{X}_2,...,\mymat{X}_L$; ${\rm{diag}}(\mymat{x})$ denotes a diagonal matrix whose main diagonal consists of the elements of the vector $\mymat{x}$.

\section{System Model}
\label{sec_system_model}

In the considered mMTC network, a BS equipped with $M$ antennas provides service for $K_{{\rm{tot}}}$ single-antenna devices concurrently. During a specific time interval, an indeterminate number $K_{\rm{a}}< K_{{\rm{tot}}}$ of devices are active, each of which transmits $B$-bits message directly to the BS without a scheduling process. 
We assume that all of the active devices transmit synchronously on the common $T$ available time-frequency resource elements \cite{p_ccs_ura_angle, p_ccs_ura_nonb, p_ccs_ura_amp}. 
The primary responsibility of the BS is to recover the set of the transmitted messages, irrespective of the individual device identities.

\subsection{Signal Model}

\begin{figure}[!t]
	\centering
	\includegraphics[width=0.25\textwidth]{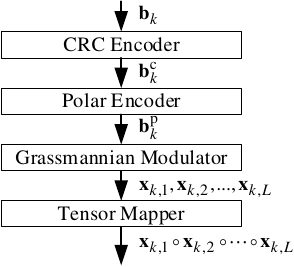}
	\caption{The block diagram of the transmitter in PTURA.}
	\label{fig_deviceside}
\end{figure}

Dentoe $\mymat{h}_{k}\in\mathbb{C}^{M\times 1}$ as the channel of active device $k$. We consider the quasi-static Rayleigh fading channel model, i.e., 
\begin{align}
	\mymat{h}_{k}\sim\mathcal{CN}(\mymat{h}_{k};\mymat{0}_M,\mymat{I}_{M}),
\end{align}
where $\mymat{0}_{M}$ denotes the all-zero vector with the length $M$, and $\mymat{I}_{M}$ denotes the identity matrix with the dimension $M$. 
Let $\{\mymat{x}_{k,l}:k=1,...,L\}$ denote the Grassmannian symbols of active device $k$, where $\mymat{x}_{k,l}\in\mathbb{C}^{T_l\times 1}$, and the length $T_l$ satisfies $\prod_{l=1}^LT_l=T$. 
Then, the tensor signal $\mathcal{Y}\in\mathbb{C}^{T_1\times T_2\times\cdots\times T_{L}\times M}$ received at the BS is given by
\begin{align}
	\mathcal{Y}&=\sum\limits_{k\in \mathcal{K}_{\rm{a}}}\mymat{x}_{k,1}\circ\mymat{x}_{k,2}\circ\cdots\circ\mymat{x}_{k,L}\circ\mymat{h}_{k}+\mathcal{Z},
	\label{equ_signal_trans}
\end{align}
where $\mathcal{K}_{\rm{a}}$ denotes the set of active devices, the cardinality $|\mathcal{K}_{\rm{a}}|=K_{\rm{a}}$ is unknown to the BS, and $\mathcal{Z}\in\mathbb{C}^{T_1\times T_2\times\cdots\times T_{L}\cdots\times M}$ is the zero-mean Gaussian-i.i.d. noise tensor with the variance being $N_0$, which is an unknown variable to the BS. 
As depicted in Fig. \ref{fig_deviceside}, $\mymat{x}_{k,1}\circ\mymat{x}_{k,2}\circ\cdots\circ\mymat{x}_{k,L}\in\mathbb{C}^{T_1\times T_2\times\cdots\times T_{L}}$ represents the tensor symbol transmitted from active device $k$ to the BS, which is obtained from the message $\mymat{b}_k\in\{0,1\}^{B \times 1}$ of active device $k$ through the CRC encoding, the Polar encoding, the Grassmannian modulation, and the tensor mapping processes.

\subsection{Proposed PTURA Scheme}
The efficacy of current practical TURA is limited due to its suboptimal coding and decoding techniques \cite{p_tb_ura}. 
We incorporate contemporary advanced coding methodologies into the TURA to fully exploit its performance potential. Specially, given the short packet transmission characteristic intrinsic to mMTC systems, we leverage the Polar coding for its superior error correction capabilities. Additionally, the PTURA employs the Grassmannian modulation, ensuring the accuracy of demodulation. As illustrated in Fig. \ref{fig_deviceside}, the PTURA transmitter encompasses the CRC encoder, Polar encoder, Grassmannian modulator, and tensor mapper.

\noindent{\it 1) CRC Encoder:}

The CRC coding, a prevalent error-detection technique in communication transmissions \cite{p_3gpp_code}, is implemented in the PTURA scheme for verifying the decoding results. In the CRC encoding process, the message $\mymat{b}_k$ is encoded to the CRC codeword $\mymat{b}^{\rm{c}}_k\in\{0,1\}^{B^{\rm{c}}\times 1}$. 

\noindent{\it 2) Polar Encoder:}

The Polar code demonstrates exceptional error-correcting capability for short messages, and has been selected as the channel code for the control information transmission in contemporary 5G communicaitons \cite{p_3gpp_code}. 
With the $(B^{\rm{p}},B^{\rm{c}})$ Polar encoder, the CRC codeword $\mymat{b}^{\rm{c}}_k$ is encoded to the Polar codeword $\mymat{b}^{\rm{p}}_k\in\{0,1\}^{B^{\rm{p}}\times 1}$.

\noindent{\it 3) Grassmannian Modulation:}

The Grassmannian constellation embeds a reference signal into each symbol to identify the subspace and eliminate the phase ambiguity of the symbol, which ensures the accuracy of the demodulation in the TURA \cite{p_tb_ura}. We employ the low-complexity structured Grassmannian modulation in \cite{p_grass}, which begins with the segmentation of the message. Based on the segment length $T_1,T_2,...,T_L$, active device $k$ partions $\mymat{b}^{\rm{p}}_k$ into $L$ parts $\{\mymat{b}_{k,l}^{\rm{p}}:l=1,...,L\}$, where $\mymat{b}_{k,l}^{\rm{p}}\in\{0,1\}^{B_l^{\rm{p}}\times 1}$, $ \sum_{l=1}^LB_l^{\rm{p}}=B^{\rm{p}}$, and $B_l^{\rm{p}}-\log_2T_l$ is proportional to $T_l-1$ to maximize the minimum distance of the Grassmannian symbols, i.e.,
\begin{align}
B_l^{\rm{p}}=\log_2T_l+\frac{B^{\rm{p}}-\log_2T}{\sum_{l=1}^LT_l-L}(T_l-1).
\end{align}

Furthermore, $\mymat{b}_{k,l}^{\rm{p}}$ is modulated as the Grassmannian sysmbol $\mymat{x}_{k,l}$.
The first $\log_2T_l$ bits of
$\mymat{b}^{\rm{p}}_{k,l}$ indicate the position of the reference signal in $\mymat{x}_{k,l}$.
The rest $B_l^{\rm{p}}-\log_2T_l$ bits of
$\mymat{b}^{\rm{p}}_{k,l}$ are uniformly divided  into $2(T_l-1)$ segments $\{\mymat{b}^{{\rm{p}}}_{k,l,g}:g=1,...,2(T_l-1)\}$, where $\textstyle{\mymat{b}^{{\rm{p}}}_{k,l,g}\in\{0,1\}^{B_l^{\rm{p}}/(2T_l-2)\times 1}}$. 
Then, with the Gray-coded pulse amplitude modulation,
$\mymat{b}^{{\rm{p}}}_{k,l,g}$ is mapped as the scalar $a_{k,l,g}$ in
$$
\left\{(2n-1)2^{-\frac{B_l^{\rm{p}}}{2(T_l-1)}-1}:n=1,...,2^{\frac{B_l^{\rm{p}}}{2(T_l-1)}}\right\}.
$$ 
Denote $\omega_{k,l,i}=F_{\rm{c}}^{-1}(a_{k,l,2i-1})+jF_{\rm{c}}^{-1}(a_{k,l,2i}),i=1,...,T_l-1,$
where $F_{\rm{c}}(\cdot)$ represents the cumulative distribution function of the standard real Gaussian distribution, and $F_{\rm{c}}^{-1}(\cdot)$ represents the inverse function of $F_{\rm{c}}(\cdot)$. Let
\begin{align}
t_{k,l,i}=\sqrt{\frac{1-\exp(-\frac{|\omega_{k,l,i}|^2}{2})}{1+\exp(-\frac{|\omega_{k,l,i}|^2}{2})}}\cdot\frac{\omega_{k,l,i}}{|\omega_{k,l,i}|},i=1,...,T_l-1.
\label{equ_grass_22}
\end{align}
Then, the Grassmannian symbol $\mymat{x}_{k,l}$ is given by
\begin{align}
\mymat{x}_{k,l}=\sqrt{\frac{T_l}{1+\sum_{i=1}^{T_l-1}|t_{k,l,i}|^2}}[t_{k,l,1},t_{k,l,2},...,1,...,t_{k,l,T_l-1}]^T,
\label{equ_grass_symbol}
\end{align}
where $\sqrt{\frac{T_l}{1+\sum_{i=1}^{T_l-1}|t_{k,l,i}|^2}}$ is the reference signal, and its position is determined by the first $\log_2T_l$ bits of
$\mymat{b}^{\rm{p}}_{k,l}$.

\noindent{\it 4) Tensor Mapper:}

After the Grassmannian modulation,
the Grassmannian symbols $\left\{\mymat{x}_{k,l}:l=1,...,L\right\}$ from device $k$ is mapped to the tensor symbol $\mymat{x}_{k,1}\circ\mymat{x}_{k,2}\circ\cdots\circ\mymat{x}_{k,L}\in\mathbb{C}^{T_1\times T_2\times \cdots\times T_L}$ to be transmitted to the BS.

\section{Iterative Bayesian Receiver with Feedback}
\label{sec_iter_receiver}
\begin{figure}[!h]
	\centering
	\includegraphics[width=0.375\textwidth]{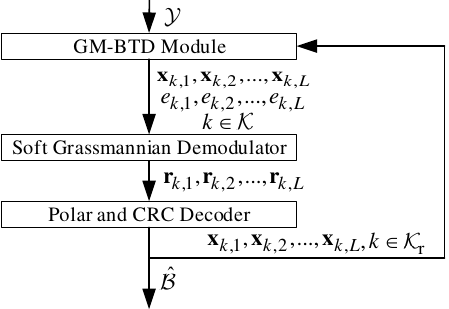}
	\caption{The block diagram of the IBR-FB.}
	\label{fig_rec_sic}
\end{figure}

Given the noisy observation $\mathcal{Y}$, the decoder at the BS
is tasked with estimating the transmitted messages set $\mathcal{B}=\left\{\mymat{b}_k:k\in\mathcal{K}_{\rm{a}}\right\}$. To this end, we propose the IBR-FB, which consists of the GM-BTD module, the soft Grassmannian demodulator, and the Polar and CRC decoder. 
As depicted in Fig. \ref{fig_rec_sic}, within each iteration of the IBR-FB, the processing procedure is as: 1) The GM-BTD generates the soft information about the unrecovered messages, utilizing the received tensor signals and the regenerated Grassmannian symbols corresponding to the currently recovered valid messages. 2) On the basis of the soft information provided by the GM-BTD, the soft Grassmannian demodulation and the Polar decoding, as well as the CRC decoding, are performed sequentially to obtain the newly recovered valid messages satisfying the CRC requirement.
3) The newly and currently recovered valid messages form the recovered valid messages for the next iteration, whose corresponding Grassmannian symbols are fed back to the GM-BTD.

\begin{remark}
The proposed IBR-FB differs from the successive interference cancellation (SIC) receiver. Within each iteration of the IBR-FB, only the regenerated Grassmannian symbols corresponding to the currently recovered valid messages are fed back, while the channels corresponding to the currently recovered valid messages continue to be unknown.
\end{remark}

The remainder of this section provides a detailed description of the processing procedure within one iteration of the IBR-FB.

\subsection{GM-BTD Module}
\label{sec_gm_btd_module}

Denote the set of the active devices that transmit the currently recovered valid messages as $\mathcal{K}_{{\rm{r}}}$, where the cardinality $|\mathcal{K}_{{\rm{r}}}|=K_{{\rm{r}}}$.
With the same encoder and modulator described in section \ref{sec_system_model}, we regenerate the Grassmannian symbols associated with the currently recovered valid messages, denoted as 
$\left\{\mymat{x}_{k,l}:k\in\mathcal{K}_{{\rm{r}}},l=1,...,L\right\}$\footnote{We disregard the false alarms in  $\left\{\mymat{x}_{k,l}:k\in\mathcal{K}_{\rm{r}},l=1,...,L\right\}$, which arises from the limited error detection capability of the CRC code.}.
Then, the signal transmission model can be reformulated as
\begin{align}
	\textstyle
	\mathcal{Y}&=\sum_{k\in\mathcal{K}_{\rm{u}}}\underbrace{\mymat{x}_{k,1}\circ\mymat{x}_{k,2}\circ\cdots\circ\mymat{x}_{k,L}}_{\text {unrecovered}}\circ\mymat{h}_{k}\nonumber\\
	&+\sum_{k\in\mathcal{K}_{\rm{r}}}\underbrace{\mymat{x}_{k,1}\circ\mymat{x}_{k,2}\circ\cdots\circ\mymat{x}_{k,L}}_{\text{currently recovered}}\circ\mymat{h}_{k}+\mathcal{Z},
	\label{equ_new_signal_trans}
\end{align}
where $\mathcal{K}_{\rm{u}}=\mathcal{K}_{{\rm{a}}}\setminus\mathcal{K}_{\rm{r}}$ represents the set of the active devices that transmit the unrecovered messages, and the cardinality $|\mathcal{K}_{{\rm{u}}}|=K_{\rm{u}}$ is unknown.
Estimating the set of the unrecovered Grassmannian symbols $\left\{\mymat{x}_{k,l}:k\in\mathcal{K}_{\rm{u}},l=1,...,L\right\}$ based on (\ref{equ_new_signal_trans}) is a TD problem.

\begin{remark}
When $\mathcal{K}_{\rm{r}}$ is empty, the TD problem degenerates into the conventional CPD problem.
\end{remark}

Enabling the soft decoding for the PTURA necessitates
the incorporation of the TD module capable of producing
soft outputs, which can not be attained by the conventional
optimization-based algorithms \cite{p_op_td1,p_op_td2}. To resolve this challenge,
we propose the GM-BTD algorithm under the variational
Bayesian learning framework \cite{p_paper_vb,p_paper_vb_2}, which can produce the required soft outputs to cooperate with the PTURA.
As shown in Fig. \ref{fig_rec_sic}, given the inputs $\mathcal{Y}$ and $\left\{\mymat{x}_{k,l}:k\in\mathcal{K}_{{\rm{r}}},l=1,...,L\right\}$,
the GM-BTD
algorithm generates the estimation of the unrecovered Grassmannian symbols $\hat{\mymat{x}}_{k,l}\in\mathbb{C}^{T_l\times 1},k=1,...,\hat K_{\rm u},l=1,...,L$, 
and the corresponding estimation error $e_{k,l},k=1,...,\hat K_{\rm{u}},l=1,...,L$, where $||\hat{\mymat{ x}}_{k,l}||_2^2=T_l$, $\hat K_{\rm{u}}$ denotes the estimation of $K_{\rm{u}}$, and
\begin{align}
e_{k,l}=\frac{\mathbb{E}\left\{||\hat{\mymat{ x}}_{k,l}-\mathbb{E}\{\hat{\mymat{x}}_{k,l}\}||_2^2\right\}}{T_l}.
\end{align}

A comprehensive description of the proposed GM-BTD algorithm will be presented in section \ref{sec_bayes_td}. 

\begin{algorithm}[!t]
	\caption{IBR-FB} 
	\label{al_pr_fb}
	{\bf {Input:}} $\mathcal{Y}$.
	
	{\bf {Initialize:}} $\mathcal{K}_{{\rm{r}}}=\emptyset$, $\mathcal{\hat B}=\emptyset$.
	\begin{algorithmic}[1]
		\Repeat
		\State 
		Execute the GM-BTD (Algorithm \ref{al_vb_td}) with the in-
		\hspace*{0.16in} puts $\mathcal{Y}$ and $\left\{\mymat{x}_{k,l}:k\in\mathcal{K}_{\rm{r}},l=1,...,L\right\}$ to export
		$\hat{\mymat{x}}_{k,l}$
		\hspace*{0.16in} and $e_{k,l}$.
		\State 
		Execute the soft Grassmannian demodulation (\ref{equ_ap_llr}) with			
		\hspace*{0.16in} the inputs $\hat{\mymat{x}}_{k,l}$ and $e_{k,l}$ to output $\mymat{r}_{k,l}$. 
		\State 
		Perform the Polar and CRC decoding to output the	
		\hspace*{0.16in} newly recovered valid messages set $\mathcal{\hat B}^{\rm{new}}$. 
		\State
		$\mathcal{\hat B}=\mathcal{\hat B}\cup\mathcal{\hat B}^{\rm{new}}$.
		\State Update $\mathcal{K}_{\rm{r}}$ and $\left\{\mymat{x}_{k,l}:k\in\mathcal{K}_{\rm{r}},l=1,...,L\right\}$.
		\Until {$\mathcal{K}_{\rm{r}}$ ceases to change.}
	\end{algorithmic}
	\hspace*{0in} {\bf {Output:}} $\mathcal{\hat B}$.	
\end{algorithm}

\subsection{Soft Grassmannian Demodulation}

On the basis of $\hat{\mymat{x}}_{k,l}$ and $e_{k,l}$, the soft Grassmannian demodulation is performed to obtain the log-likelihood ratio $\mymat{r}_{k,l}\in\mathbb{R}^{B_l^{\rm{p}}\times 1}$ associated with the unrecovered messages, where the entry $\mymat{r}_{k,l}[i]$ is given by
\begin{align}
\mymat{r}_{k,l}[i]=\ln\frac{\sum_{\mymat{x}_{k,l}\in\mathcal{M}_{k,l,i}^1  } p(\hat{\mymat{x}}_{k,l}|\mymat{x}_{k,l},e_{k,l})}{ \sum_{\mymat{x}_{k,l}\in\mathcal{M}_{k,l,i}^0 }p(\hat{\mymat{x}}_{k,l}|\mymat{x}_{k,l},e_{k,l})},
\end{align} 
where the set $\mathcal{M}_{k,l,i}^{q}=\{\mymat{x}_{k,l}:\mymat{b}_{k,l}^{\rm{p}}[i]=q\},q=0,1,$ includes all the Grassmannian symbols corresponding to the Polar codewords, for which the $i$-th bits are equal to $q$. 
Building upon the foundations established in \cite{p_tb_bound_ura}, an approximated closed-form log-likelihood ratio is given by
\begin{align}
	\mymat{r}_{k,l}[i]
	& \approx\frac{2\sqrt{1-e_{k,l}}}{e_{k,l}}\left(\left|\hat{\mymat{ x}}_{k,l}^H\mymat{x}_{k,l,i}^1\right|-\left|\hat{\mymat{ x}}_{k,l}^H\mymat{x}_{k,l,i}^0\right|\right),
	\label{equ_ap_llr}
\end{align}
where the Grassmannian symbol $\mymat{x}_{k,l,i}^{q},q=0,1,$ is defined as
\begin{align}
	\mymat{x}_{k,l,i}^{q}={\rm{arg}}\mathop{\rm { max}}_{\mymat{x}_{k,l}\in\mathcal{M}_{k,l,i}^{q}}\,\,\left|\mymat{ x}_{k,l}^H\hat{\mymat{x}}_{k,l}\right|.
\end{align}

It is worth noting that $\mymat{x}_{k,l,i}^{q}$ corresponds to the result of the maximum likelihood (ML) Grassmannian demodulation:
\begin{align}
	{\mymat{b}}_{k,l}^{{\rm{p,ML}}}={\rm{arg}}\mathop{\rm { max}}_{\mymat{b}_{k,l}^{\rm{p}}\in\{0,1\}^{B_l^{\rm{p}}\times 1}}\,\,\left|\mymat{ x}_{k,l}^H\hat{\mymat{x}}_{k,l}\right|,
	\label{equ_ml_grass}
\end{align}
where $\mymat{x}_{k,l}$ represents the Grassmannian symbol corresponding to $\mymat{b}_{k,l}^{\rm{p}}$. Denote $\mymat{x}_{k,l}^{{\rm{ML}}}$ as the Grassmannian symbol corresponding to $\mymat{b}_{k,l}^{{\rm{p,ML}}}$. 
If the $i$-th bit of $\mymat{b}_{k,l}^{{\rm{p,ML}}}$ is equal to 1, then $\mymat{x}_{k,l,i}^{1}=\mymat{x}_{k,l}^{{\rm{ML}}}$, and $\mymat{x}_{k,l,i}^{0}$ is the nearest Grassmannian symbol to $\mymat{x}_{k,l}^{{\rm{ML}}}$ with the $i$-th bit in its corresponding Polar codeword being 0. Otherwise,  $\mymat{x}_{k,l,i}^{0}=\mymat{x}_{k,l}^{{\rm{ML}}}$, and $\mymat{x}_{k,l,i}^{1}$ is the nearest Grassmannian symbol to $\mymat{x}_{k,l}^{{\rm{ML}}}$ with the $i$-th bit in its corresponding Polar codeword being 1.
$\mymat{x}_{k,l}^{{\rm{ML}}}$ is computed through the low-complexity greedy decoder proposed in \cite{p_grass}\footnote{It has been demonstrated that the greedy decoder with the computational complexity being $2(T_l-1)2^{{B_l^{\rm{p}}}/{(2T_l-2)}}$ can achieve near-ML performance.}, based on which $\mymat{x}_{k,l,i}^{q}$ is obtained. Following this, the log-likelihood ratio $\mymat{r}_{k,l}$ is directly determined by (\ref{equ_ap_llr}).

\subsection{Polar and CRC Decoder}
The Polar decoding and the CRC decoding are performed with the CRC-aided SCL Polar decoder, where the Polar decoder provides a list of $n_{{\rm{list}}}$ candidate messages, and the CRC decoder verifies these messages  \cite{p_scl,p_scl_crc}. 
With the log-likelihood ratio $\mymat{r}_{k}=[\mymat{r}_{k,1}^T,\mymat{r}_{k,2}^T,...,\mymat{r}_{k,L}^T]^T\in\mathbb{R}^{B^{\rm{p}}\times 1}$, the CRC-aided SCL Polar decodor yields the newly recovered valid messages $\mathcal{\hat B}^{\rm{new}}$ satisfying the CRC requirement.

The newly and currently recovered valid messages form the recovered valid messages of the next iteration, whose corresponding Grassmannian symbols will be fed back to the GM-BTD. Such iterative process continues until the set  $\mathcal{K}_{\rm{r}}$ ceases to change.
Finally, the recovered valid messages in the last iteration constitute the estimation $\mathcal{\hat B}$ of the transmitted messages set $\mathcal{B}$. We summarize the proposed IBR-FB in Algorithm \ref{al_pr_fb}.

\section{The Grassmannian Modulation-Aided Bayesian TD Algorithm}
\label{sec_bayes_td}
We propose the GM-BTD algorithm under the
variational Bayesian learning framework to generate
the required soft information, leveraging the property of the Grassmannian modulation to facilitate the convergence of the variational Bayesian learning process.
With the inputs $\mathcal{Y}$ and $\left\{\mymat{x}_{k,l}:k\in\mathcal{K}_{{\rm{r}}},l=1,...,L\right\}$,
the GM-BTD outputs the estimation $\hat K_{\rm{u}}$ of the unrecovered messages number, the estimation $\hat{\mymat{ x}}_{k,l}$ of the unrecovered Grassmannian symbols,
and the correspoding estimation error $e_{k,l}$, without the need for the number $K_{\rm{a}}$ of active devices.

Due to the intricate relationship between the number of the unrecovered messages $K_{\rm{u}}$ and the received tensor signal $\mathcal{Y}$, 
directly estimating $K_{\rm{u}}$ poses a significant challenge \cite{p_vb_cpd,p_gvb_cpd}.  
To address this issue, we consider a predefined set $\mathcal{K}$ with the cardinality $|\mathcal{K}|=K$ ($\mathcal{K}_{\rm u}\subset\mathcal{K}$), which is composed of $K_{\rm {u}}$ active devices transmitting the unrecovered messages and $K-K_{\rm {u}}$ inactive devices\footnote{The details for ascertaining $K$ are provided in section \ref{sec_simulation}. Note that $K-K_{\rm {u}}$ represents the number of inactive devices introduced artificially, not the actual number of inactive devices in the network.}. Then, we can obtain the estimation of $K_{\rm{u}}$ by determining the number of inactive devices in $\mathcal{K}$, thus aiding in the construction of the probabilistic model and further facilitating the design of the TD algorithm.

With the set $\mathcal{K}$, we can express the signal transmission model (\ref{equ_new_signal_trans}) more compactly as
\begin{align}
	&\mathcal{Y}=\sum\limits_{k\in\mathcal{K}}\underbrace{\mymat{x}_{k,1}\circ\mymat{x}_{k,2}\circ\cdots\circ\mymat{x}_{k,L}}_{\text{unrecovered or all-zero}}\circ\mymat{h}_k\nonumber\\
	&+\sum\limits_{k\in\mathcal{K}_{\rm{r}}}\underbrace{\mymat{x}_{k,1}\circ\mymat{x}_{k,2}\circ\cdots\circ\mymat{x}_{k,L}}_{\text{currently recovered}}\circ\mymat{h}_k+\mathcal{Z}\nonumber\\
	&=\left[\left|\mymat{X}_{\mathcal{K},1},...,\mymat{X}_{\mathcal{K},L},\mymat{H}_{\mathcal{K}}\right|\right]+\left[\left|\mymat{X}_{\mathcal{K}_{\rm{r}},1},...,\mymat{X}_{\mathcal{K}_{\rm{r}},L},\mymat{H}_{\mathcal{K}_{\rm{r}}}\right|\right]+\mathcal{Z},
	\label{equ_signal_trans2}
\end{align}
where the symbol (channel) factor matrix $\mymat{X}_{\mathcal{K},l}\in\mathbb{C}^{T_l\times K}$ ($\mymat{H}_{\mathcal{K}}\in\mathbb{C}^{M\times K}$) associated with the set $\mathcal{K}$ consists of an all-zero submatrix with column of $K-K_{\rm{u}}$ and another submatrix that comprises $K_{\rm{u}}$ Grassmannian symbols (channels) corresponding to the unrecovered messages, the symbol (channel) factor matrix $\mymat{X}_{\mathcal{K}_{\rm{r}},l}\in\mathbb{C}^{T_l\times K_{{\rm{r}}}}$ ($\mymat{H}_{\mathcal{K}_{\rm{r}}}\in\mathbb{C}^{M\times K_{{\rm{r}}}}$) associated with the set $\mathcal{K}_{\rm{r}}$ consists of $K_{\rm{r}}$ regenerated Grassmannian symbols (channels) corresponding to the currently recovered valid messages, and $\left[\left|\cdots\right|\right]$ represents the Kruskal operator \cite{p_op_td2}.

\subsection{Probabilistic Model}
\label{sec_vb_pf}

According to the signal transmission model, we can get the likelihood probability distribution as shown in (\ref{equ_lr_dis}).
\begin{figure*}[!t]
	\centering
	\begin{align}
		p\left(\mathcal{Y}|\{\mymat{X}_{\mathcal{K},l}:l=1,...,L\},\mymat{H}_{\mathcal{K}},\mymat{H}_{\mathcal{K}_{\rm{r}}},N_0\right)=\prod_{i_1,...,i_L,m}\mathcal{CN}\left(\mathcal{Y}[i_1,...,i_L,m];
		\sum_{k\in\{\mathcal{K},\mathcal{K}_{\rm{r}}\}}\mymat{h}_k[m]\prod_{l=1}^{L}\mymat{x}_{k,l}[i_l],N_0\right)
		\label{equ_lr_dis}
	\end{align}
	\noindent\rule{\textwidth}{0.5pt}\nonumber
\end{figure*}
We model the unknown noise precision $N_0^{-1}$ with the Gamma distribution \cite{p_paper_vb}, i.e.,
\begin{align}
p\left(\frac{1}{N_0}\right)={\rm{G}}\left(\frac{1}{N_0};a_0,b_0\right)=\frac{b_0^{a_0}N_0^{1-a_0}e^{-b_0N_0^{-1}}}{\Gamma(a_0)},
\end{align} 
where $\Gamma(\cdot)$ represents the Gamma function. As no prior information about the noise can be acquired by the BS, we set the hyperparameters $a_0\to 0$ and $b_0\to 0$, resulting in an uninformative prior ${\rm{G}}(N_0^{-1};a_0,b_0)$.

The prior of the symbol factor matrix $\mymat{X}_{\mathcal{K},l}$ is characterized by a complex discrete distribution associated with the Grassmannian modulation constellation, combined with the inherent sparsity. In order to simplify the joint probability distribution and develop a tractable inference algorithm, we approximate this discrete distribution as a continuous Gaussian distribution, and employ a Gamma distributed precision to capture the inherent sparsity, i.e.,
\begin{align}
p(\mymat{X}_{\mathcal{K},l}|\mymat{\lambda})&=\prod_{k\in \mathcal{K}}\mathcal{CN}(\mymat{x}_{k,l};\mymat{0}_{T_l},\lambda_k^{-1}\mymat{I}_{T_l}),
\end{align} 
where $\mymat{\lambda}=[\lambda_1,\lambda_2,...,\lambda_K]^T\in\mathbb{R}^{K\times 1}$, and the precision $\lambda_k$ exhibits the Gamma distribution, i.e., 
\begin{align}
p(\mymat{\lambda})=\prod_{k\in\mathcal{K}}{\rm{G}}(\lambda_{k}; a_{{\rm{\lambda}},k}, b_{{\lambda},k}).
\end{align}
The sparsity of $\mymat{X}_{\mathcal{K},l}$ is controled by $\mymat{\lambda}$. For instance,
when $\lambda_{k}\to \infty$, the $k$-th column of $\mymat{X}_{\mathcal{K},l}$ approaches $\mymat{0}_{T_l}$. 
It should be noted that the precision $\mymat{\lambda}$ is shared by $L$ symbol factor matrices to facilitate capturing the joint sparsity.
Moreover, since the BS cannot obtain prior information regarding the number of unrecovered messages, we set the hyperparameters $a_{{\lambda},k}\to 0$ and $b_{{\lambda},k}\to 0$. 

As for the channel factor matrix $\mymat{H}_{\mathcal{K}}$, we adopt an uninformative Gamma distributed precision to capture its inherent sparsity, i.e., 
\begin{align}
&p(\mymat{H}_{\mathcal{K}}|\mymat{\gamma})=\prod_{k\in{\mathcal{K}}}\mathcal{CN}(\mymat{h}_{k};\mymat{0}_M,\gamma_k^{-1}\mymat{I}_M),
\end{align}
where $\mymat{\gamma}=[\gamma_1,\gamma_2,...,\gamma_K]^T\in\mathbb{R}^{K\times 1}$, and the precision $\gamma_k$ exhibits the Gamma distribution, i.e., 
\begin{align}
p(\mymat{\gamma})=\prod_{k\in\mathcal{K}}{\rm{G}}(\gamma_{k}; a_{{{\rm{\gamma}}},k}, b_{{{\rm{\gamma}}},k}),
\end{align}
where we set the hyperparameters $a_{{{\rm{\gamma}}},k}\to 0$ and $b_{{{\rm{\gamma}}},k}\to 0$.

Since the channel factor matrix $\mymat{H}_{\mathcal{K}_{\rm{r}}}$ does not exhibit the sparsity characteristic, it is prior probability distributioncan be written as 
\begin{align}
	p(\mymat{H}_{\mathcal{K}_{\rm{r}}})=\prod_{k\in\mathcal{K}_{\rm{r}}}\mathcal{CN}(\mymat{h}_{k};\mymat{0}_{M},\mymat{I}_{M}).
\end{align}
 
\begin{figure*}[!t]
	\begin{align}
		p(\mathcal{P}|\mathcal{Y})\propto p\left(\mathcal{Y}|\{\mymat{X}_{\mathcal{K},l}:l=1,...,L\},\mymat{H}_{\mathcal{K}},\mymat{H}_{\mathcal{K}_{\rm{r}}},N_0\right)p(N_0^{-1}) p(\mymat{H}_{\mathcal{K}_{\rm{r}}})p(\mymat{H}_{\mathcal{K}}|\mymat{\gamma})p(\mymat{\gamma})p(\mymat{\lambda}) \prod_{l=1}^{L}p(\mymat{X}_{\mathcal{K},l}|\mymat{\lambda})
		\label{equ_joint_pro}
	\end{align}
\noindent\rule{\textwidth}{0.5pt}\nonumber
\end{figure*}

Then, the joint posterior probability distribution of the system can be expressed as (\ref{equ_joint_pro}),
where the unknown random variables set $\mathcal{P}=\left\{\{\mymat{X}_{\mathcal{K},l}:l=1,...,L\},\mymat{H}_{\mathcal{K}},\mymat{H}_{\mathcal{K}_{\rm{r}}},\mymat{\lambda},\mymat{\gamma},N_0^{-1}\right\}$. On the basis of (\ref{equ_joint_pro}), the TD problem described in section \ref{sec_gm_btd_module} can be solved by performing Bayesian estimation (e.g., MMSE estimation) for $\mathcal{P}$.

\subsection{Variational Bayesian Learning}
\label{sec_vb_learning}

Exact Bayesian inference based on (\ref{equ_joint_pro}) necessitates multiple integrations, rendering the process intractable.
To takle of this challenge, we resort to the variational Bayesian learning method. 
In particular, we seek for a tractable distribution $q^\star(\mathcal{P})$ to approximate the ground truth joint posterior probability distribution by minimizing the Kullback-Leibler (KL) divergence, i.e.,
\begin{align}
q^\star(\mathcal{P})={\rm{arg}}\,\mathop{\rm{min}}_{q(\mathcal{P})}\,\, \int q(\mathcal{P})\ln\frac{q(\mathcal{P})}{p(\mathcal{P}|\mathcal{Y})} {\rm d}\mathcal{P}.
\label{equ_real_op}
\end{align}
Subsequently, based on the distribution $q^\star(\mathcal{P})$, the unknown random variables set $\mathcal{P}$ is estimated using the MMSE criterion, which is given by
\begin{align}
\mathcal{\hat P}&=\mathbb{E}_{q^\star(\mathcal{P})}\{\mathcal{P}\}.
\label{equ_vb_total}
\end{align}

The KL divergence in (\ref{equ_real_op}) achieves the minimum value when $q(\mathcal{P})=p(\mathcal{P}|\mathcal{Y})$. In the absence of any constraints imposed on $q(\mathcal{P})$, $q^\star(\mathcal{P})=p(\mathcal{P}|\mathcal{Y})$, which reverts us to the original intractable posterior probability distribution.
A prevalent approach to address this issue is the mean field approximation \cite{p_book_graph}. In this context, 
$q(\mathcal{P})$ can be factorized as
\begin{align}
q(\mathcal{P})
&=q(N_0^{-1})q(\mymat{\lambda})q(\mymat{\gamma})q(\mymat{H}_{\mathcal{K}})q(\mymat{H}_{\mathcal{K}_{\rm{r}}})\prod_{l=1}^{L}q(\mymat{X}_{\mathcal{K},l}).
\label{equ_ass}
\end{align}
Furthermore, the factorized form in (\ref{equ_ass}) implies the adoption of the iterative block coordinate descent methodology, capable of acquiring a suboptimal solution of (\ref{equ_real_op}). Denote $q(\mathcal{P}_i)$ as the $i$-th factor in $q(\mathcal{P})$. By fixing the factors $\{q(\mathcal{P}_j):j\neq i\}$, the optimum form of the factor $q(\mathcal{P}_i)$ is given by 
\begin{align}
\ln q(\mathcal{P}_i)=\mathbb{E}_{q(\mathcal{P}\setminus \mathcal{P}_i)}\{\ln p(\mathcal{Y},\mathcal{P})\}+\text{const}.
\label{equ_vbu}
\end{align}
Leveraging (\ref{equ_vbu}), $q^\star(\mathcal{P})$ is obtained by updating each factor $q(\mathcal{P}_i)$ until achieving the convergence. In the remainder of this section, we derive the closed-form iterative rules for each factor.

\subsubsection{{The  Symbol Factor $ q(\mathbf{X}_{\mathcal{K},l}) $}}
By applying (\ref{equ_vbu}), we can get (\ref{equ_first_com}),
\begin{figure*}[!t]
	\centering
\begin{align}
\ln q(\mymat{X}_{\mathcal{K},l})&=-\frac{1}{\hat{N}_0}{\rm{tr}}\left(\mymat{X}_{\mathcal{K},l}^T\mymat{X}_{\mathcal{K},l}^*\left[\mymat{J}_{l}\circledast\left(\hat{\mymat{ H}}_{\mathcal{K}}^H\hat{\mymat{ H}}_{\mathcal{K}}+\sum_{m=1}^{M}\mymat{\Phi}_{m}^*\right)\right]\right)
-{\rm{tr}}\left(\mymat{X}_{\mathcal{K},l}^T\mymat{X}_{\mathcal{K},l}^*{\rm{diag}}(\hat{\mymat{\lambda}})\right)\nonumber\\
&+\frac{2}{\hat{N}_0}\Re\left\{{\rm{tr}}\left(\mathcal{Y}_{(l)}^T\mymat{X}_{\mathcal{K},l}^*\left(\hat{\mymat{ H}}_{\mathcal{K}}\odot\mymat{V}_{l}\right)^H-\mathcal{W}_{(l)}^T\mymat{X}_{\mathcal{K},l}^*\left(\hat{\mymat{ H}}_{\mathcal{K}}\odot\mymat{V}_{l}\right)^H\right)\right\}+\text{const}
	\label{equ_first_com}
\end{align}
\noindent\rule{\textwidth}{0.5pt}\nonumber
\end{figure*}
which indicates that $q(\mymat{X}_{\mathcal{K},l})$ is a Gaussian distribution, different rows in $\mymat{X}_{\mathcal{K},l}$ are independent of each other, the posterior mean $\hat{\mymat{ X}}_{\mathcal{K},l}[i,:]=\mathbb{E}_{q(\mathcal{P})}\{\mymat{X}_{\mathcal{K},l}[i,:]\}$ and the corresponding covariance matrix $\mymat{\Theta}_{i,l}=\mathbb{E}_{q(\mathcal{P})}\left\{\mymat{X}_{\mathcal{K},l}[i,:]^T\mymat{X}_{\mathcal{K},l}[i,:]^*\right\}-{\hat{\mymat{ X}}_{\mathcal{K},l}}[i,:]^T\hat{\mymat{ X}}_{\mathcal{K},l}[i,:]^*$ are given by 
\begin{align}
	&\mymat{\Theta}_{i,l}=\left(\frac{1}{\hat{N}_0}\left[\mymat{J}_{l}\circledast\left(\hat{\mymat{ H}}_{\mathcal{K}}^H\hat{\mymat{ H}}_{\mathcal{K}}+\sum_{m=1}^{M}\mymat{\mymat{\Phi}}_{m}^*\right)\right]+{\rm{diag}(\hat{\mymat{ \lambda}})}\right)^{-1},\label{equ_v_x}\\
	&\hat{\mymat{ X}}_{{\mathcal{K}},l}[i,:]=\frac{1}{\hat{N}_0}\left(\mathcal{Y}_{(l)}[i,:]-\mathcal{W}_{(l)}[i,:]\right)\left(\hat{\mymat{ H}}_{\mathcal{K}}\odot \mymat{V}_{l}\right)^*\mymat{\Theta}_{i,l}^T,\label{equ_e_x}
\end{align} 
where $\hat{N}_0^{-1}=\mathbb{E}_{q(\mathcal{P})}\{N_0^{-1}\}$ represents the posterior mean of $N_0^{-1}$, $\mymat{J}_{l}\in\mathbb{C}^{K\times K}$ is defined as
\begin{align}
\mymat{J}_{l}\defequ{\mathlarger{\mathlarger{{\mathlarger{\circledast}}}}}_{l'=1,l'\neq l}^L\left(\hat{\mymat{ X}}_{\mathcal{K},l'}^H\hat{\mymat{ X}}_{\mathcal{K},l'}+\sum_{i=1}^{T_{l'}}\mymat{\Theta}_{i,l'}^*\right),
\label{equ_jjl}
\end{align}
$\hat{\mymat{ H}}_{\mathcal{K}}=\mathbb{E}_{q(\mathcal{P})}\{\mymat{H}_{\mathcal{K}}\}$ represents the posterior mean of $\mymat{H}_{\mathcal{K}}$, $\mymat{\Phi}_{m}=\mathbb{E}_{q(\mathcal{P})}\left\{\mymat{H}_{\mathcal{K}}[m,:]^T\mymat{H}_{\mathcal{K}}[m.:]^*\right\}-\hat{\mymat{H}}_{\mathcal{K}}[m,:]^T\hat{\mymat{H}}_{\mathcal{K}}[m,:]^*$
represents the posterior covariance matrix of $\mymat{H}_{\mathcal{K}}[m,:]$, $\hat{\mymat{ \lambda}}=\mathbb{E}_{q(\mathcal{P})}\{\mymat{\lambda}\}$ represents the posterior mean of $\mymat{\lambda}$, $\mymat{V}_{l}\in\mathbb{C}^{\frac{T}{T_l}\times K}$ is defined as
\begin{align}
\mymat{V}_{l}\defequ\hat{\mymat{ X}}_{\mathcal{K},L}\odot\cdots\odot \hat{\mymat{ X}}_{\mathcal{K},l+1}\odot\hat{\mymat{ X}}_{\mathcal{K},l-1}\odot\cdots\odot \hat{\mymat{ X}}_{\mathcal{K},1},
\label{equ_vvl}
\end{align}
$\mathcal{Y}_{(l)}\in\mathbb{C}^{T_l\times \frac{TM}{T_l}}$ and $\mathcal{W}_{(l)}\in\mathbb{C}^{T_l\times \frac{TM}{T_l}}$ represents the mode-$l$ expansion of $\mathcal{Y}$ and
\begin{align}
\mathcal{W}=\left[\left|\mymat{X}_{\mathcal{K}_{\rm{r}},1},...,\mymat{X}_{\mathcal{K}_{\rm{r}},L},\hat{\mymat{ H}}_{\mathcal{K}_{\rm{r}}}\right|\right],
\end{align}
$\hat{\mymat{ H}}_{\mathcal{K}_{\rm{r}}}=\mathbb{E}_{q(\mathcal{P})}\left\{\mymat{H}_{\mathcal{K}_{\rm{r}}}\right\}$ represents the posterior mean of $\mymat{H}_{\mathcal{K}_{\rm{r}}}$. 
It is important to highlight that the covariance matrix $\mymat{\Theta}_{i,l}$ is dependent on the mode index $l$, not the row index $i$, implying that only a single covariance matrix $\mymat{\Theta}_{l}$ requires computation for the $T_l$ rows of $\mymat{X}_{\mathcal{K},l}$.

\begin{figure*}[!t]
	\begin{align}
		\ln q(\mymat{H}_{\mathcal{K}})=-\frac{1}{\hat{N}_0}{\rm{tr}}\left(\mymat{H}_{\mathcal{K}}^T\mymat{H}_{\mathcal{K}}^*\mymat{J}\right)-{\rm{tr}}\left(\mymat{H}_{\mathcal{K}}^T\mymat{H}_{\mathcal{K}}^*{\rm{diag}}(\hat{\mymat{\gamma}})\right)
		+\frac{2}{\hat{N}_0}\Re\left\{{\rm{tr}}\left(\left(\mathcal{Y}_{(L+1)}-\mathcal{W}_{(L+1)}\right)^T\mymat{H}_{\mathcal{K}}^*\mymat{V}^H\right)\right\}+\text{const}\label{equ_complex_2}
	\end{align}
\noindent\rule{\textwidth}{0.5pt}\nonumber
\end{figure*}

\subsubsection{{The  Channel Factor $ q(\mathbf{H}_{\mathcal{K}}) $}}
By applying (\ref{equ_vbu}), we can get (\ref{equ_complex_2}),
where $\mymat{J}\in\mathbb{C}^{K\times K}$ is defined as
\begin{align}
	\mymat{J}\defequ{\mathlarger{\mathlarger{{\mathlarger{\circledast}}}}}_{l=1}^L\left(\hat{\mymat{ X}}_{\mathcal{K},l}^H\hat{\mymat{ X}}_{\mathcal{K},l}+T_l\mymat{\Theta}_{l}^*\right),
	\label{equ_jj}
\end{align}
$\hat{\mymat{ \gamma}}=\mathbb{E}_{q(\mathcal{P})}\{\hat{\mymat{\gamma}}\}$ denotes the posterior mean of $\mymat{\gamma}$, and $\mymat{V}\in\mathbb{C}^{T\times K}$ is defined as
\begin{align}
	\mymat{V}\defequ\hat{\mymat{ X}}_{\mathcal{K},L}\odot\hat{\mymat{ X}}_{\mathcal{K},L-1}\odot\cdots\odot\hat{\mymat{ X}}_{\mathcal{K},1}.
	\label{equ_vv}
\end{align}
Hence $q(\mymat{H}_{\mathcal{K}})$ is a Gaussian distribution, different rows in $\mymat{H}_{\mathcal{K}}$ are independent of each other, the posterior mean $\hat{\mymat{ H}}_{\mathcal{K}}[m,:]$ and the corresponding covariance matrix $\mymat{\Phi}_{m}$ are given by 
\begin{align}
\mymat{\Phi}_m&=\left(\frac{1}{\hat{N}_0}\mymat{J}+{\rm{diag}(\hat{\mymat{ \gamma}})}\right)^{-1},\label{equ_v_h}\\
\hat{\mymat{ H}}_{\mathcal{K}}[m,:]&=\frac{1}{\hat{N}_0}\left(\mathcal{Y}_{(L+1)}[m,:]-\mathcal{W}_{(L+1)}[m,:]\right)\mymat{V}^*\mymat{\Phi}_m^T.\label{equ_e_h}
\end{align}
Notably, the covariance matrix $\mymat{\Phi}_m$ is not influenced by the row index $m$, meaning that computing a single covariance matrix $\mymat{\Phi}$ is sufficient for the $M$ rows of $\mymat{H}_{\mathcal{K}}$.

\begin{figure*}[!t]
	\begin{align}
		\ln q(\mymat{H}_{\mathcal{K}_{\rm{r}}})&=-\frac{1}{\hat{N}_0}{\rm{tr}}\left(\mymat{H}_{\mathcal{K}_{\rm{r}}}^T\mymat{H}_{\mathcal{K}_{\rm{r}}}^*\left[\circledast_{l=1}^L\left(\mymat{X}_{\mathcal{K}_{\rm{r}},l}^H\mymat{ X}_{\mathcal{K}_{\rm{r}},l}\right)\right]\right)-{\rm{tr}}\left(\mymat{H}_{\mathcal{K}_{\rm{r}}}^T\mymat{H}_{\mathcal{K}_{\rm{r}}}^*\right)\nonumber\\
		&+\frac{2}{\hat{N}_0}\Re\left\{{\rm{tr}}\left(\mathcal{Y}_{(L+1)}^T\mymat{H}_{\mathcal{K}_{\rm{r}}}^*\left(\odot_{l={L}}^1\mymat{X}_{\mathcal{K}_{\rm{r}},l}\right)^H-\mathcal{X}_{(L+1)}^T\mymat{H}_{\mathcal{K}_{\rm{r}}}^*\left(\odot_{l={L}}^1\mymat{X}_{\mathcal{K}_{\rm{r}},l}\right)^H\right)\right\}+\text{const}
		\label{equ_complex_3}
	\end{align}
	\noindent\rule{\textwidth}{0.5pt}\nonumber
\end{figure*}

\subsubsection{{The  Channel Factor $ q(\mathbf{H}_{\mathcal{K}_{\rm{r}}}) $}}
For the channel factor $ q(\mathbf{H}_{\mathcal{K}_{\rm{r}}})$, we can get (\ref{equ_complex_3}),
where $\mathcal{X}\in\mathbb{C}^{T_1\times T_2\times\cdots\times T_L\times M}$ is given by
\begin{align}
\mathcal{X}=[|\hat{\mymat{ X}}_{\mathcal{K},1},...,\hat{\mymat{ X}}_{\mathcal{K},L},\hat{\mymat{ H}}_{\mathcal{K}}|].
\end{align}
Therefore, $q(\mymat{H}_{\mathcal{K}_{\rm{r}}})$ is a Gaussian distribution, different rows in $\mymat{H}_{\mathcal{K}_{\rm{r}}}$ are independent of each other, and the posterior mean $\hat{\mymat{H}}_{\mathcal{K}_{\rm{r}}}[m,:]$ is given by
\begin{align}
	\hat{\mymat{ H}}_{\mathcal{K}_{\rm{r}}}[m,:]&=\frac{1}{\hat N_0}\left(\mathcal{Y}_{(L+1)}[m,:]-\mathcal{X}_{(L+1)}[m,:]\right)\nonumber\\
	&\times \left(\odot_{l={L}}^1\mymat{ X}_{\mathcal{K}_{\rm{r}},l}\right)^*\mymat{\Xi}^T,\label{equ_e_s}
\end{align}
where $\mymat{\Xi}\in\mathbb{C}^{K_{\rm{r}}\times K_{\rm{r}}}$ is given by
\begin{align}
\mymat{\Xi}&=\left(\frac{1}{\hat{N}_0}\left[\circledast_{l=1}^L\left(\mymat{X }_{{\mathcal{K}_{\rm{r}}},l}^H\mymat{ X}_{\mathcal{K}_{\rm{r}},l}\right)\right]+\mymat{I}_{K_{{\rm{r}}}}\right)^{-1}\label{equ_v_s}.
\end{align}

\subsubsection{{The Factor $q(N_0^{-1})$}}
By applying (\ref{equ_vbu}), the posterior probability distribution $q(N_0^{-1})$ satisfies (\ref{equ_qp_n}),
\begin{figure*}[!ht]
\begin{align}
&\ln q(N_0^{-1})=-(TM+a_0-1)\ln N_0
-N_0^{-1}\left(b_0+||\mathcal{Y}||_F^2+\mymat{1}_{K}^T\left[\mymat{J}\circledast(\hat{\mymat{ H}}_{\mathcal{K}}^H\hat{\mymat{ H}}_{\mathcal{K}}
	+M\mymat{\Phi}^*)\right]\mymat{1}_{K}\right.\nonumber\\
	&\left.+\mymat{1}_{K_{{\rm{r}}}}^T\left[\left(\circledast_{l=1}^{L}\left(\mymat{X}_{\mathcal{K}_{\rm{r}},l}^H\mymat{ X}_{\mathcal{K}_{\rm{r}},l}\right)\right)\circledast\left(\hat{\mymat{ H}}_{\mathcal{K}_{\rm{r}}}^H\mymat{ \hat H}_{\mathcal{K}_{\rm{r}}}+M\mymat{\Xi}\right)\right]\mymat{1}_{K_{{\rm{r}}}}-2\Re\left\{\left<\mathcal{Y},\mathcal{W}+\mathcal{X}\right>\right\}+2\Re\left\{\left<\mathcal{W},\mathcal{X}\right>\right\}\right)+\text{const}
	\label{equ_qp_n}
\end{align}
\noindent\rule{\textwidth}{0.5pt}\nonumber
\end{figure*}
where
\begin{align}
||\mathcal{Y}||_F^2=\sum_{i_2,i_2,...,i_L,m}\left|\mathcal{Y}[i_1,i_2,...,i_L,m]\right|^2,
\end{align}
$\mymat{1}_K$ denotes an all-one vector with the length $K$, and $<\mathcal{W},\mathcal{X}>$ denotes the inner product of $\mathcal{W}$ and $\mathcal{X}$.
Then, we can conclude $q(N_0^{-1})$ is a Gamma distribution, i.e., $q(N_0^{-1})={\rm{G}}(N_0^{-1};c_0,d_0)$, where
\begin{align}
	&d_0=b_0+||\mathcal{Y}||_F^2+\mymat{1}_{K}^T\left[\mymat{J}\circledast(\hat{\mymat{ H}}_{\mathcal{K}}^H\hat{\mymat{ H}}_{\mathcal{K}}
	+M\mymat{\Phi}^*)\right]\mymat{1}_{K}\nonumber\\
	&+\mymat{1}_{K_{{\rm{r}}}}^T\left[\left(\circledast_{l=1}^{L}\left(\mymat{X}_{\mathcal{K}_{\rm{r}},l}^H\mymat{ X}_{\mathcal{K}_{\rm{r}},l}\right)\right)\circledast\left(\hat{\mymat{ H}}_{\mathcal{K}_{\rm{r}}}^H\mymat{ \hat H}_{\mathcal{K}_{\rm{r}}}+M\mymat{\Xi}\right)\right]\mymat{1}_{K_{{\rm{r}}}}\nonumber\\
	&-2\Re\left\{\left<\mathcal{Y},\mathcal{W}+\mathcal{X}\right>\right\}+2\Re\left\{\left<\mathcal{W},\mathcal{X}\right>\right\},\\
	&c_0=TM+a_0.
\end{align} 
Furthermore, the posterior mean $\hat{N}_0^{-1}$ of the noise precision is given by
\begin{align}
\hat{N}_0^{-1}=\frac{c_0}{d_0}.\label{equ_e_n}
\end{align} 

\subsubsection{{The Factor \texorpdfstring{$q(\mymat{\lambda})$}{}}}
For the random vector $\mymat{\lambda}$, its posterior probability distribution $q(\mymat{\lambda})$ satisfies
\begin{align}
\ln q(\mymat{\lambda})&=\text{const}+\sum_{k\in\mathcal{K}}\left(\left(\sum_{l=1}^LT_l+a_{{\lambda},k}-1\right)\ln \lambda_{k}\right.\nonumber\\
&\left.-\left(\sum_{l=1}^L\left(\hat{\mymat{ x}}_{k,l}^H\hat{\mymat{ x}}_{k,l}+T_l\mymat{\Theta}_{l}[k,k]\right)+b_{{\lambda},k}\right)\lambda_{k}\right),
\end{align}
where $\hat{\mymat{ x}}_{k,l}$ represents the $k$-th column of $\hat{\mymat{ X}}_{\mathcal{K}}$. Therefore, $q(\mymat{\lambda})$ is a Gamma distribution, and the  
posterior mean of $\lambda_{k}$ is given by
\begin{align}
	\hat{\lambda}_{k}=\frac{\sum_{l=1}^LT_l+a_{{\lambda},k}}{\sum_{l=1}^L\left(\hat{\mymat{ x}}_{k,l}^H\hat{\mymat{ x}}_{k,l}+T_l\mymat{\Theta}_{l}[k,k]\right)+b_{{\lambda},k}}.\label{equ_e_lambda}
\end{align}

\subsubsection{{The Factor \texorpdfstring{$q(\mymat{\gamma})$}{}}}
For the random vector $\mymat{\gamma}$, its posterior probability distribution $q(\mymat{\gamma})$ satisfies
\begin{align}
	\ln q(\mymat{\gamma})&=\text{const}+\sum_{k\in\mathcal{K}}\left(\left(M+a_{{\rm{\gamma}},k}-1\right)\ln \gamma_{k}\right.\nonumber\\
	&\left.-\left(\hat{\mymat{ h}}_{k}^H\hat{\mymat{ h}}_{k}+M\mymat{\Phi}[k,k]+b_{\gamma,k}\right)\gamma_{k}\right),
\end{align}
where $\hat{\mymat{ h}}_{k}$ represents the $k$-th column of $\hat{\mymat{ H}}_{\mathcal{K}}$.
The 
posterior mean $\hat{\gamma}_{k}$ is given by
\begin{align}
\hat{\gamma}_{k}=\frac{M+a_{\gamma,k}}{\hat{\mymat{ h}}_{k}^H\hat{\mymat{ h}}_{k}+M\mymat{\Phi}[k,k]+b_{\gamma,k}}.\label{equ_e_lambda1}
\end{align}

It is worth noting that the iterative rules derived above are based on the assumption that $\mymat{x}_{k,l},k\in\mathcal{K}$, exhibits the Gaussian-Gamma prior probability distribution, where the Grassmannian modulation is not considered. We suggest that the constant energy property of the Grassmannian symbols, i.e., $||\mymat{x}_{k,l}||_2^2=T_l$, can be utilized to make the factor matrix $\hat{\mymat{ X}}_{\mathcal{K},l}$ converge towards the ground truth factor matrix $\mymat{X}_{\mathcal{K},l}$.
Dentoe $\mymat{M}_l={\rm{diag}}\left(\frac{\sqrt{T_l}}{||\hat{\mymat{ x}}_{1,l}||_2},\frac{\sqrt{T_l}}{||\hat{\mymat{ x}}_{2,l}||_2},...,\frac{\sqrt{T_l}}{||\hat{\mymat{ x}}_{K,l}||_2}\right)$ as the modified matrix. After a round of updates in the variational Bayesian learning process, we revise the posterior mean and the posterior covariance matrix of $\mymat{X}_{\mathcal{K},l}$ as
\begin{align}
\hat{\mymat{ X}}_{\mathcal{K},l}&\leftarrow\hat{\mymat{ X}}_{\mathcal{K},l}\mymat{M}_l,\quad l=1,...,L,\label{equ_revb}\\
\hat{\mymat{H}}_{\mathcal{K}}&\leftarrow\hat{\mymat{ H}}_{\mathcal{K}}\left(\prod_{l=1}^L\mymat{M}_l\right)^{-1},\\
{\mymat{\Theta}}_l&\leftarrow\mymat{M}_l\mymat{\Theta}_l\mymat{M}_l^H,\quad l=1,...,L,\\
{\mymat{\Phi}}&\leftarrow\left(\prod_{l=1}^L\mymat{M}_l\right)^{-1}\mymat{\Phi}\left(\prod_{l=1}^L\mymat{M}_l\right)^{-H}.
\label{equ_reve}
\end{align}
where $\leftarrow$ represents the assignment operation.

The number of non-zero columns of $\mymat{X}_{\mathcal{K},l}$ or $\mymat{H}_{\mathcal{K}}$ in (\ref{equ_signal_trans2}) can be interpretated as the number $ K_{\rm{u}}$ of the unrecovered messages. After the revisions (\ref{equ_revb})--(\ref{equ_reve}), the estimation $\hat K_{\rm{u}}$ is represented by the number of non-zero columns of $\hat{\mymat{ H}}_{\mathcal{K}}$. We obtain $\hat K_{\rm {u}}$ by the energy detection method, i.e., the $k$-th column in $\hat{\mymat{ H}}_{\mathcal{K}}$ is considered as non-zero only if
\begin{align}
||\hat{\mymat{ h}}_{k}||^2_2>\epsilon_{\rm{a}},
\label{equ_dim_reduce}
\end{align}
where $\epsilon_{\rm{a}}>0$ represents a predefined threshold.
Furthermore, in order to expedite the variational Bayesian learning process, the 
on-the-fly pruning is employed \cite{p_vb_cpd}. Specifically, during each iteration, if the energy of any column in $\hat{\mymat{ H}}_{\mathcal{K}}$ drops below the predefined threshold $\epsilon_{\rm{a}}$, the corresponding device is pruned from $\mathcal{K}$. Consequently, $\mathcal{K}$ diminishes with the number of iterations, and $\hat K_{\rm u}$ is the cardinality of $\mathcal{K}$ upon convergence of the variational Bayesian learning process.

We summarize the proposed GM-BTD algorithm in Algorithm \ref{al_vb_td}. 

\begin{algorithm}[!t]
	\caption{GM-BTD Algorithm} 
	\label{al_vb_td}
	{\bf {Input:}} $\mathcal{Y}$, $\epsilon_{\rm{a}}$, $\epsilon_{{\rm{iter}}}$, $\{\mymat{X}_{\mathcal{K}_{\rm{r}},l}:\,\forall l\}$.\\
	{\bf {Initialize:}} $\{\hat{\mymat{ X}}_{\mathcal{K},l},\mymat{\Theta}_l:\,\forall l\}$, $\{a_{\gamma,k},b_{\gamma,k},a_{\lambda,k}, b_{\lambda,k}:\,\forall k\}$, $\hat{\mymat{ H}}_{\mathcal{K}}$, $\mymat{\Phi}$, $\hat{\mymat{ H}}_{\mathcal{K}_{\rm{r}}}$, $\mymat{\Xi}$, $a_0$,
	$b_0$.
	\begin{algorithmic}[1]
		\State 
		$\hat{N}_0^{-1}=a_0/b_0$.
		\State
		$\hat{\lambda}_{k}=a_{\lambda,k}/b_{\lambda,k}, \forall k$.
		\State
		$\hat{\gamma}_{k}=a_{\gamma,k}/b_{\gamma,k}, \forall k$.
		\State
		$\hat{\mymat{ X}}_{\mathcal{K},l}^{{\rm{old}}}=\hat{\mymat{ X}}_{\mathcal{K},l}, \forall l$.
		\Repeat
		\State
		Compute $\mathcal{W}=[|\mymat{X}_{\mathcal{K}_r,1},...,\mymat{X}_{\mathcal{K}_r,L},\hat{\mymat{ H}}_{\mathcal{K}_r}|]$.
		\State 
		Compute $\mymat{\Theta}_l, \forall l,$ and $\hat{\mymat{ X}}_{\mathcal{K},l}, \forall l,$ via (\ref{equ_v_x}) and (\ref{equ_e_x}). \label{line_al_1}
		\State
		Compute $\mymat{\Phi}$ and $\hat{\mymat{ H}}_{\mathcal{K}}$ via (\ref{equ_v_h}) and (\ref{equ_e_h}).\label{line_al_4}
		\State
		Compute $\hat{N}_0^{-1}$ via (\ref{equ_e_n}).
		\State
		Compute $\hat{\mymat{ \lambda}}$ via (\ref{equ_e_lambda}).
		\State
		Compute $\hat{\mymat{ \gamma}}$ via (\ref{equ_e_lambda1}).
		\State
		Combine with the Grassmannian modulation via		
		\hspace*{0.2in}(\ref{equ_revb})--(\ref{equ_reve}).
		\State
		Prune devices via (\ref{equ_dim_reduce}).
		\State
		$\hat{\mymat{ X}}_{\mathcal{K},l}^{\rm{old}}=\hat{\mymat{ X}}_{\mathcal{K},l}, \forall l$.
		\State
		Compute $\mathcal{X}=[|\hat{\mymat{ X}}_{\mathcal{K},1},...,\hat{\mymat{ X}}_{\mathcal{K},L},\hat{\mymat{ H}}_{\mathcal{K}}|]$.
		\State
		Compute $\hat{\mymat{ H}}_{\mathcal{K}_{\rm{r}}}$ via (\ref{equ_e_s}). \label{line_al_6}
		\Until{$\frac{\sum_{l=1}^{L}||\hat{\mymat{ X}}_{\mathcal{K},l}^{{\rm{old}}}-\hat{\mymat{ X}}_{\mathcal{K},l}||_F^2}{\sum_{l=1}^{L}||\hat{\mymat{ X}}_{\mathcal{K},l}^{{\rm{old}}}||_F^2}<\epsilon_{{\rm{iter}}}$}
	\end{algorithmic}
	\hspace*{0in} {\bf {Output:}} $\hat{\mymat{ X}}_{\mathcal{K},l},\forall l$ and $e_{k,l}=\mymat{\Theta}_l[k,k], \forall k,\forall l$.	
\end{algorithm}

\subsection{Computational Complexity and Convergence Property}

The computational complexity of the GM-BTD algorithm is assessed in terms of the number of complex multiplication operations. For each iteration, the computational complexity associated with calculating the relevant variables is summarized in Tab. \ref{tab_com}. With the proposed pruning method (\ref{equ_dim_reduce}), $K$ will be reduced to $K_{\rm{u}}=K_{\rm{a}}-K_{\rm{r}}$ within a few iterations (See section \ref{sec_simulation}), such that the computational complexity of the GM-BTD algorithm is of the order $\mathcal{O}(K_{\rm u}^3)$.

In each iteration, after fixing other variational distributions, the problem of optimizing a single variational distribution has been proven to be convex and the results derived in section \ref{sec_vb_learning} is its unique solution \cite{p_book_graph}. 
Each updating step in the variatioanl Bayesian learning process can be viewed as a block coordinate descent step over the functional space. The variational Bayesian learning process is guaranteed to converge to at least a stationary point of the KL divergence.

\begin{table}[!t]
	\centering 
	\caption{Computational Complexity of GM-BTD}
	\label{tab_com}
	\begin{tabular}{|c|c|}\hline
		Variable& Computational Complexity  \\ \hline
		$\mathcal{W}$& $K_{\rm{r}}TM$\\ \hline
		$\{\mymat{\Theta}_l:\forall l\}$& $K^2\sum_{l=1}^LT_l+L(L-1)K^2+LK^3$\\ \hline
		$\{\hat{\mymat{X}}_{\mathcal{K},l}:\forall l\}$& $K^2\sum_{l=1}^LT_l+LKTM$\\ \hline
		$\mymat{\Phi}$& $K^2M+(L-1)K^2+K^3$\\ \hline
		$\hat{\mymat{H}}_{\mathcal{K}}$& $K^2M+KTM$\\ \hline
		$\hat{N}_{0}^{-1}$& $LK^2+K^2_{\rm{r}}+2TM$\\ \hline
		$\hat{\mymat{\lambda}}$& $K\sum_{l=1}^LT_l$\\ \hline
		$\hat{\mymat{\gamma}}$& $KM$\\ \hline
		$\mathcal{X}$& $KTM$\\ \hline
		$\hat{\mymat{H}}_{\mathcal{K}_{\rm{r}}}$& $K_{\rm{r}}^2M+K_{\rm{r}}TM$\\ \hline
	\end{tabular}
	
\end{table}

\section{Numerical Results}
\label{sec_simulation}

\subsection{Simulation Configurations}

We present extensive simulations to validate the effectiveness of the proposed PTURA scheme under the commonly used URA settings: $T=3200$, $B=96$ and $M=50$ \cite{p_ccs_ura_nonb,p_fas_ura, p_tb_ura}. In the PTURA, we employ the CRC encoder and the Polar encoder compliant with the 5G standard \cite{p_3gpp_code}.
Various configurations of the PTURA are detailed in Tab. \ref{tab_para_tura}\footnote{Parameter selection is performed empirically here. Future investigations will focus on optimizing these parameters to augment the performance.}. 
Within each iteration of the IBR-FB, we adopt the cardinality $K$ of the considered set $\mathcal{K}$ as
\begin{align}
	K=\frac{c_K}{TM}\sum_{i_1,i_2,...,i_L,m}\left|\mathcal{Y}[i_1,i_2,...,i_L,m]-\bar y\right|^2-K_{\rm{r}},
\end{align}
where the parameter $c_K>1$ is determined by Mente Carlo simulations \footnote{It is worth noting that $K\approx c_K(K_{\rm{a}}+N_0)-K_{\rm{r}}$.
	Through Mente Carlo simulations, we ascertain that the $c_K=1.1$ sufficiently ensures $K>K_{\rm{u}}$ (i.e., $\mathcal{K}_{\rm{u}}\subset\mathcal{K}$) without resulting in excessive model intricacy.},
and $\bar y=\frac{1}{TM}\sum_{i_1,i_2,...,i_L,m}\mathcal{Y}[i_1,i_2,...,i_L,m]$ denotes the mean of $\mathcal{Y}$. Besides, the hyperparameters in the GM-BTD algorithm are set as: $a_0=b_0=10^{-6}$, $a_{\lambda,k}=b_{\lambda,k}=10^{-6},\forall k$, $a_{\gamma,k}=b_{\gamma,k}=10^{-6},\forall k$, the threshold $\epsilon_{\rm{a}}=10^{-2}$, and the iteration termination condition $\epsilon_{{\rm{iter}}}=10^{-6}$.  

\begin{figure}[!t]
\centering
\includegraphics[width=0.47\textwidth]{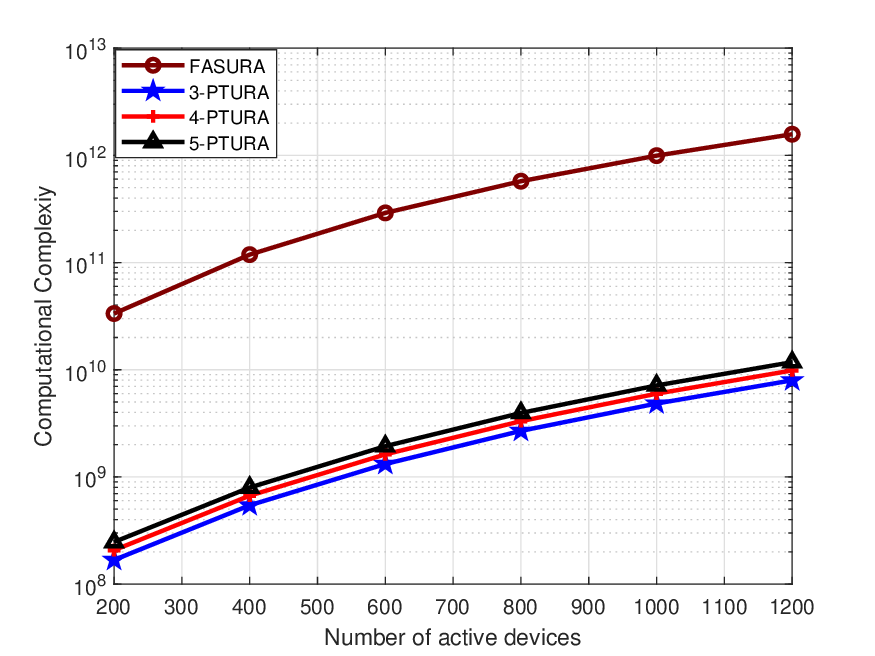}
\caption{Computational complexity in terms of required complex multiplication operations.}
\label{fig_com}
\end{figure}

With these configurations, typical values of the computational complexity corresponding to the PTURA and the sate-of-the art URA scheme, FASURA \cite{p_fas_ura}, are provided in Fig. \ref{fig_com}, where we exclude the SIC process of the FASURA and the feedback process of the PTURA for simplification. Upon observation, the PTURA
demonstrates a significantly reduced computational complexity in comparison to the FASURA.

\begin{table*}[!ht]
\centering 
\caption{Parameters of PTURA}
\label{tab_para_tura}
\begin{tabular}{|c|c|c|c|c|c|c|}\hline
Scheme& $L$ & $B^{\rm{c}}$ & $B^{\rm{p}}$ & $n_{{\rm{list}}}$ & $T_l$ & $B^{\rm{p}}_l$ \\ \hline
3-PTURA&  3 &  107  &  126  &    8        &$[T_1,T_2,T_3]=[20,16,10]$& $[B_1^{\rm{p}},B_2^{\rm{p}},B_3^{\rm{p}}]=[55,44,27]$ \\ \hline
4-PTURA&  4 &  107  &  126  &    8        & $[T_1,T_2,T_3,T_4]=[10,8,8,5]$& $[B_1^{\rm{p}},B_2^{\rm{p}},B_3^{\rm{p}},B_4^{\rm{p}}]=[42,33,33,18]$ \\ \hline
5-PTURA&  5 &  107  &  126  &    8        & $[T_1,T_2,T_3,T_4,T_5]=[8,5,5,4,4]$& $[B_1^{\rm{p}},B_2^{\rm{p}},B_3^{\rm{p}},B_4^{\rm{p}},B_5^{\rm{p}}]=[42,24,24,18,18]$ \\ \hline
\end{tabular}

\end{table*}

\begin{figure}[!t]
	\centering
	\begin{subfigure}[b]{0.94\linewidth}
		\includegraphics[width=\linewidth]{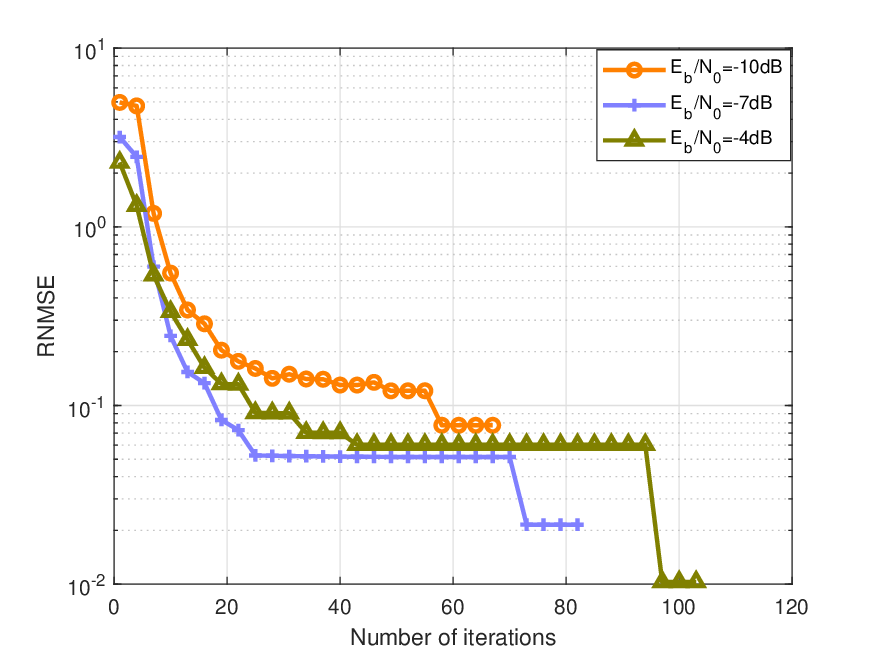}
		\caption{RNMSE}
	\end{subfigure}
	\begin{subfigure}[b]{0.94\linewidth}
		\includegraphics[width=\linewidth]{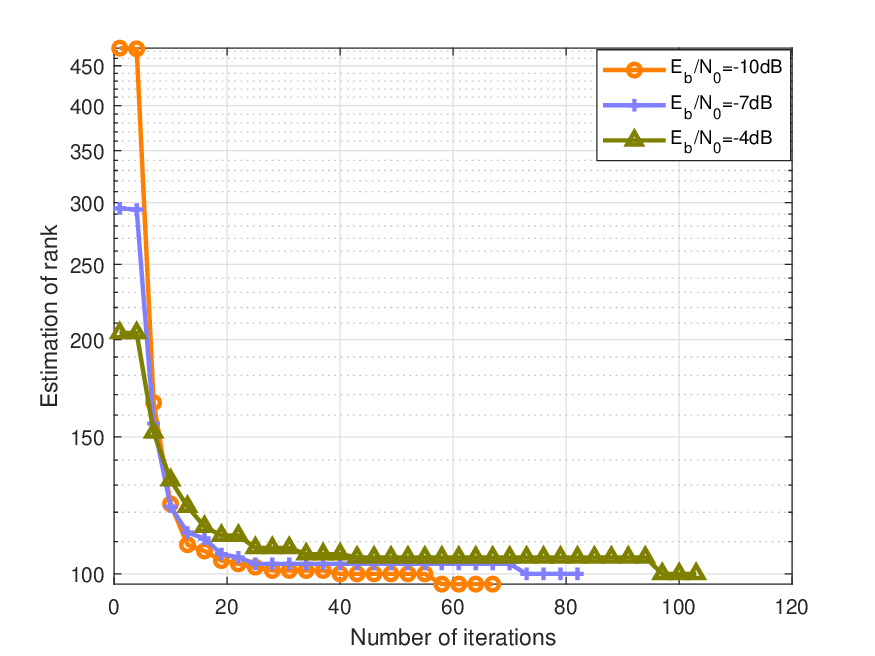}
		\caption{Estimation of Rank}
	\end{subfigure}
	\caption{The convergence behavior of the GM-BTD, where $L=3$ and $K_{\rm{a}}=100$.}
	\label{fig_sim_cpd_conver}
\end{figure}

\subsection{Performance Metrics}
In this paper, the energy per bit to noise power spectral density ratio of the system is given by $\frac{E_{\rm{b}}}{N_0}=\frac{T}{BN_0}$. Denote the collection of the unrecovered Grassmannian symbols of all $L$ segments as $\mymat{G}=[\mymat{X}_{\mathcal{K}_{\rm{u}},1}^T,...,\mymat{X}_{\mathcal{K}_{\rm{u}},L}^T]^T=[\mymat{g}_1,...,\mymat{g}_{K_{\rm{u}}}]\in\mathbb{C}^{\sum_{l=1}^LT_l\times K_{\rm{u}}}$, where $\mymat{X}_{\mathcal{K}_{\rm{u}},l}\in\mathbb{C}^{T_l\times K_{\rm{u}}}$ consists of the $K_{\rm{u}}$ unrecovered Grassmannian symbols of segment-$l$. Let $\hat{\mymat{ G}}=[\hat{\mymat{ X}}_{\mathcal{K},1}^T,...,\hat{\mymat{ X}}_{\mathcal{K},L}^T]^T=[\hat{\mymat{ g}}_1,...,\hat{\mymat{ g}}_{\hat K_{\rm{u}}}]\in\mathbb{C}^{\sum_{l=1}^LT_l\times \hat K_{\rm{u}}}$.
As the dimensions of $\mymat{G}$ and $\hat{\mymat{ G}}$ may not be the same,
we evaluate the TD performance of the GM-BTD in terms of the revised normalized mean square error (RNMSE)\footnote{We have eliminated the phase ambiguities in $\hat{\mymat{ X}}_{\mathcal{K},l}$ via the reference signals.}
\begin{align}
	{\rm{RNMSE}}= \left\{ \begin{array}{ll}
		\frac{\sum_{k=1}^{\hat K_{\rm{u}}}\mathop{\rm{min}}_{i=1,...,K_{\rm{u}}}||\hat{\mymat{ g}}_k-\mymat{g}_i||_2^2}{||\mymat{G}||_F^2},&\hat K_{\rm{u}}\ge K_{\rm{u}},\\
		\frac{\sum_{k=1}^{ K_{\rm{u}}}\mathop{\rm{min}}_{i=1,...,\hat K_{\rm{u}}}||{\mymat{ g}}_k-\hat{\mymat{g}}_i||_2^2}{||\mymat{G}||_F^2},& K_{\rm{u}}>\hat K_{\rm{u}},
	\end{array} \right.
\end{align}
and the rank estimation error ratio (REER)
\begin{align}
	{\rm{REER}}=\frac{|\hat K_{\rm{u}}-K_{\rm{u}}|}{K_{\rm{u}}}.
\end{align}
Besides, the system performance is evaluated in terms of the per-user probability of error (PUPE)
\begin{align}
	{\rm{PUPE}}=\frac{|\mathcal{B}\setminus(\mathcal{\hat B}\cap\mathcal
		{B})|}{|\mathcal{B}|}+\frac{|\mathcal{\hat B}\setminus(\mathcal{\hat B}\cap\mathcal
		{B})|}{|\mathcal{\hat B}|},
\end{align}
and the required $\frac{E_{\rm{b}}}{N_0}$
when $\text{PUPE}=0.05$.

\begin{figure*}[!t] 
	\centering	
	\begin{subfigure}[b]{0.35\textwidth} 
		\includegraphics[width=\textwidth]{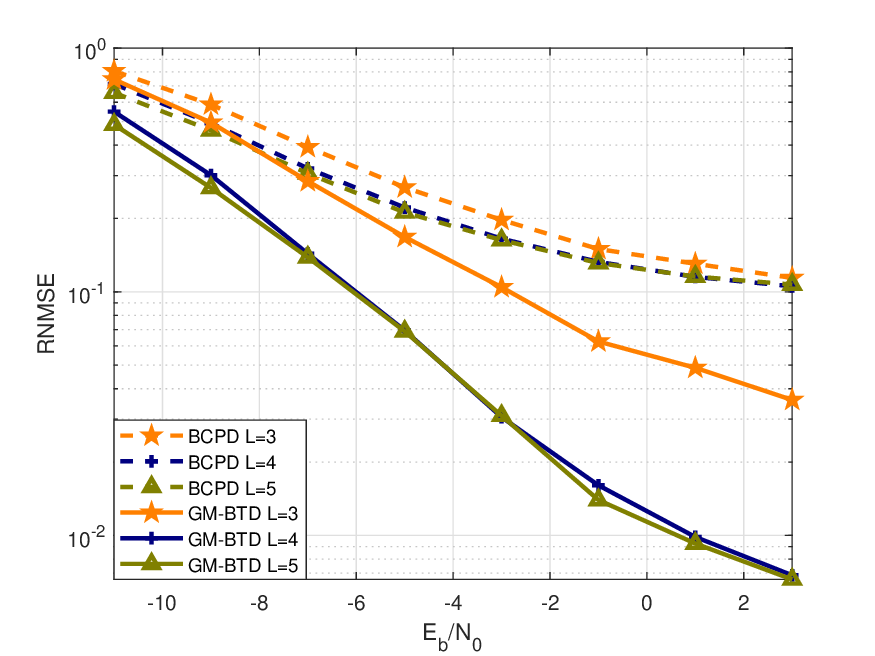}
		\caption{RNMSE: $K_{\rm{a}}=400$}
	\end{subfigure}
	\hspace{-5ex}
	\begin{subfigure}[b]{0.35\textwidth}
		\includegraphics[width=\textwidth]{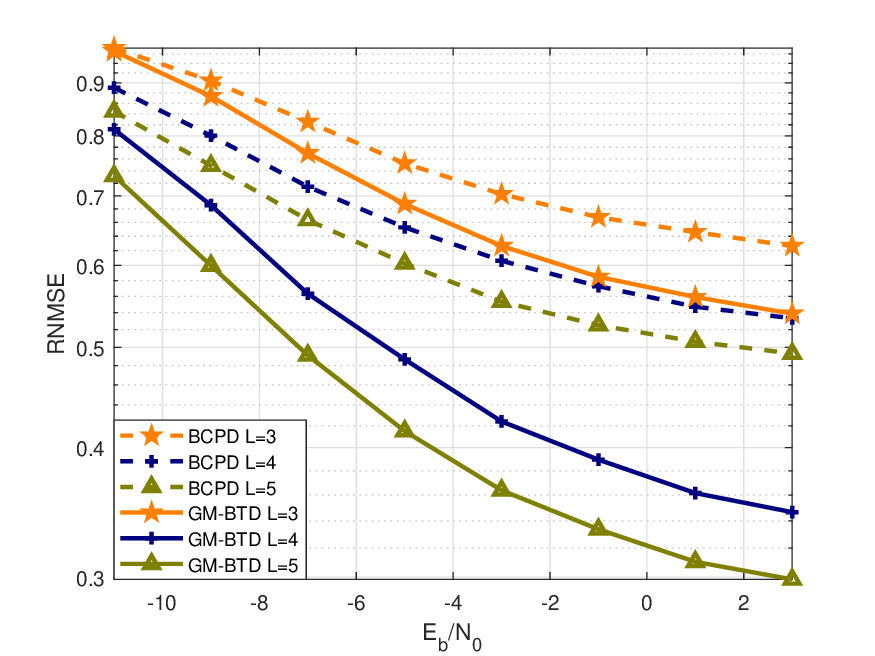}
		\caption{RNMSE: $K_{\rm{a}}=700$}
	\end{subfigure}
	\hspace{-5ex}
	\begin{subfigure}[b]{0.35\textwidth}
		\includegraphics[width=\textwidth]{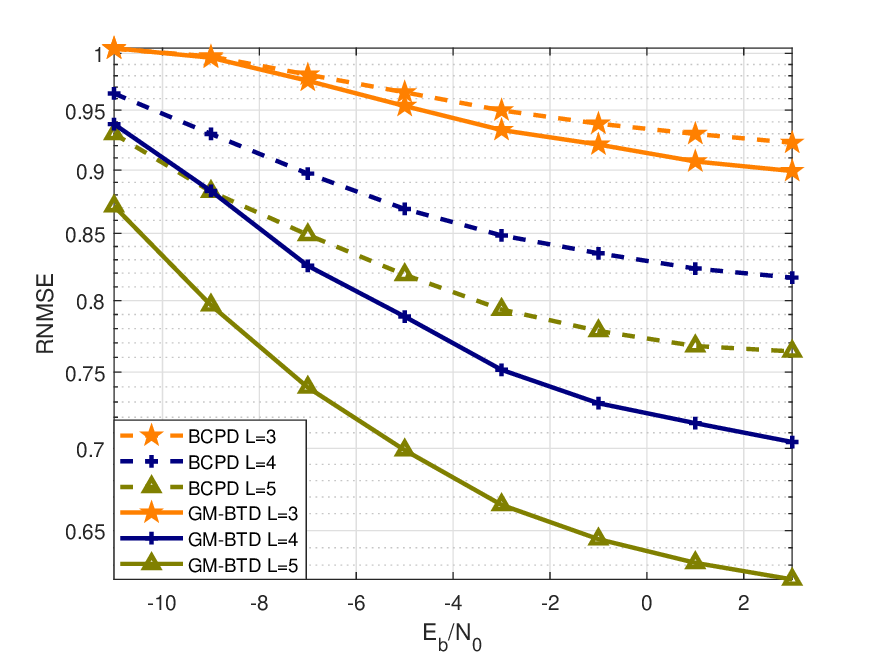}
		\caption{RNMSE: $K_{\rm{a}}=1000$}
	\end{subfigure}
	
	\begin{subfigure}[b]{0.35\textwidth} 
		\includegraphics[width=\textwidth]{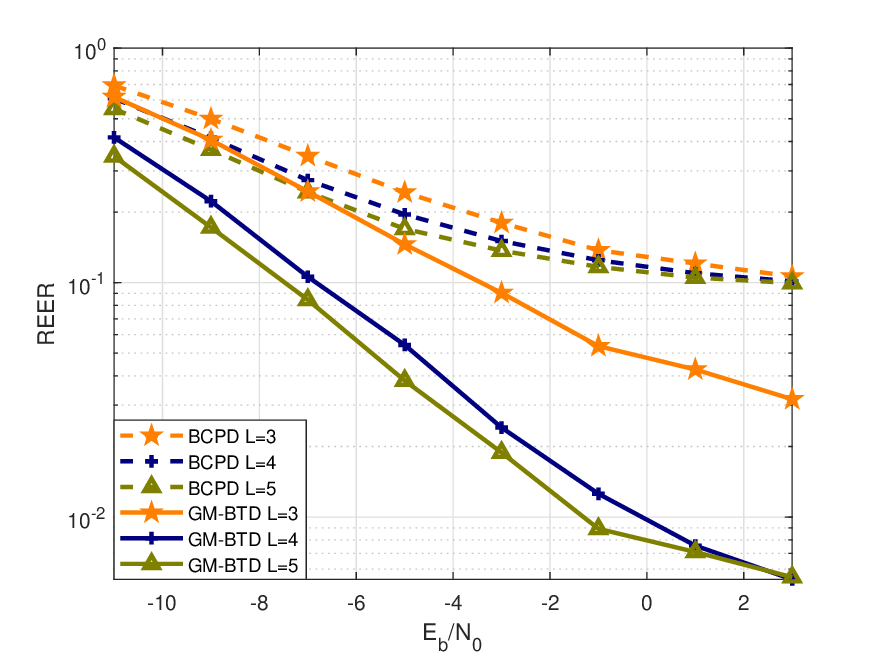}
		\caption{REER: $K_{\rm{a}}=400$}
	\end{subfigure}
	\hspace{-5ex}
	\begin{subfigure}[b]{0.35\textwidth}
		\includegraphics[width=\textwidth]{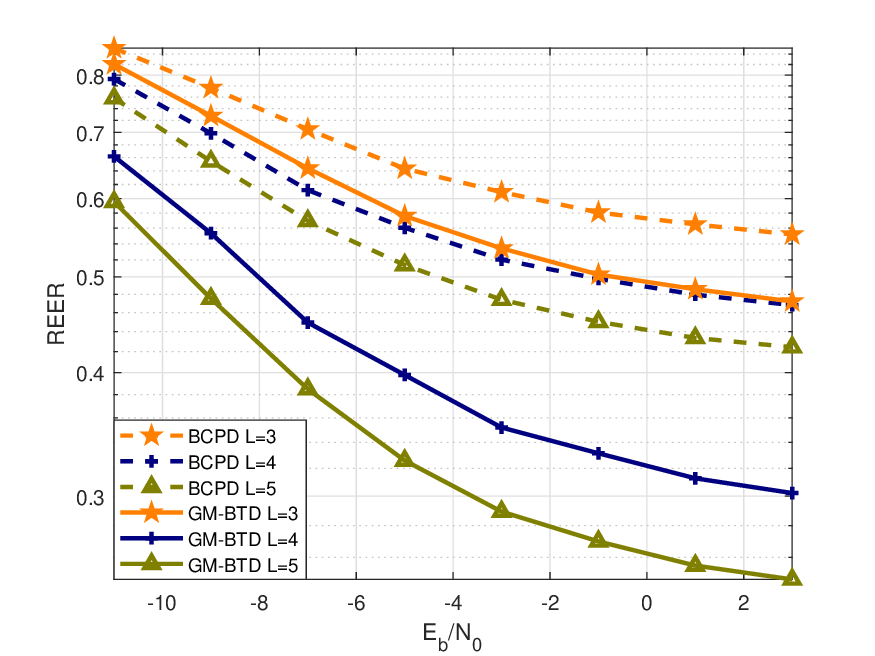}
		\caption{REER: $K_{\rm{a}}=700$}
	\end{subfigure}
	\hspace{-5ex}
	\begin{subfigure}[b]{0.35\textwidth}
		\includegraphics[width=\textwidth]{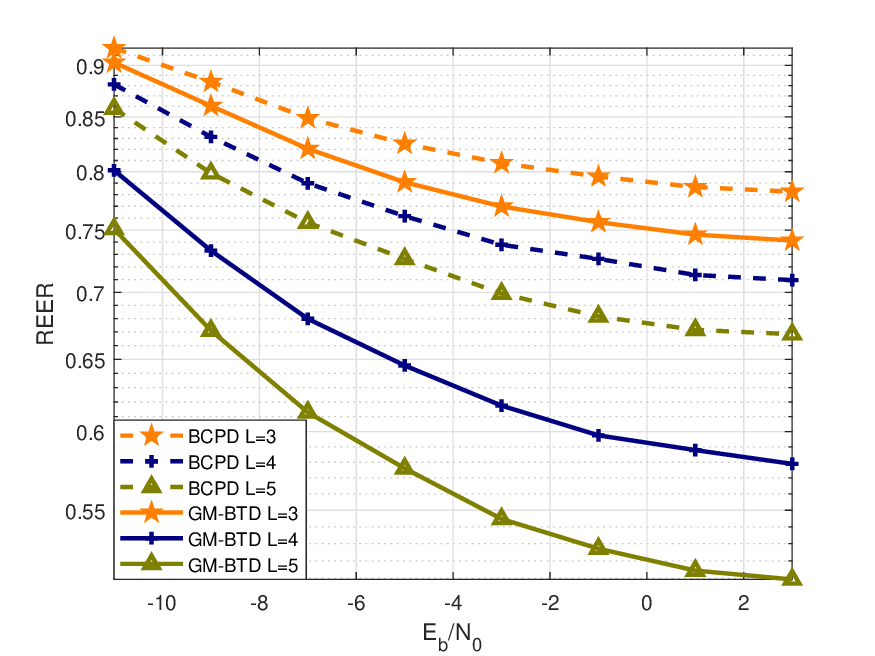}
		\caption{REER: $K_{\rm{a}}=1000$}
	\end{subfigure}
	\caption{The CPD performance comparison between the GM-BTD algorithm and the BCPD algorithm.}	
	\label{fig_sim_cpd1}
\end{figure*}

\subsection{Evaluation of the GM-BTD Module}
In this section, we consider the case where $\mathcal{K}_{\rm{r}}=\emptyset$ and $K_{\rm{u}}=K_{\rm{a}}$, causing the GM-BTD to degenerate into a CPD algorithm. A comparison is made between the GM-BTD and the Bayesian CPD algorithm (BCPD) proposed in \cite{p_vb_cpd}.
The GM-BTD algorithm and the BCPD algorithm utilize identical initializations, pruning criteria, and iteration termination condition. This ensures that the primary distinction between the GM-BTD and the BCPD algorithm lies in the consideration of Grassmannian modulation within the GM-BTD algorithm.  

Fig. \ref{fig_sim_cpd_conver} investigates the convergence behavior of the GM-BTD algorithm. The points of sharp decrease observed in Fig. \ref{fig_sim_cpd_conver} correspond to the instances where the pruning operations are executed. It can be observed that, utilizing the dimension reduction criteria (\ref{equ_dim_reduce}), the value of the initialized number of unrecovered messages $K$ in the GM-BTD will decrease approximately to $K_{\rm{a}}$, i.e., $K_{\rm{u}}$, within a few iterations. 
Fig. \ref{fig_sim_cpd1} depicts the CPD performance of the GM-BTD algorithm and the BCPD algorithm.
With the fixed block length $T$, the more number of segments $L$, the less of the unknown variables need to be estimated in the CPD process, and the larger rank of the supported tensor that can be uniquely decomposed \cite{p_an_algorithm,p_onge}. 
Such that the CPD performance of both the GM-BTD and the BCPD improve with an augmenting segment number $L$, and this phenomena accentuates increasingly as the number of the unrecovered messages $K_{\rm{u}}$ increases. 
Additionally, both the GM-BTD and the BCPD exhibit a degradation in the CPD performance with the escalation of the noise power and the proliferation of the unrecovered messages.
As the revisions (\ref{equ_revb})--(\ref{equ_reve}) in the GM-BTD enable the variational Bayesian learning process to converge towards the ground truth value, the CPD performance of the GM-BTD algorithm surpasses that of the BCPD algorithm, especially when faced with high noise power and a large number of unrecovered messages.

\subsection{Evaluation of the Polar and CRC Codes}

\begin{figure}[!t]
	\centering 
	\includegraphics[width=0.47\textwidth]{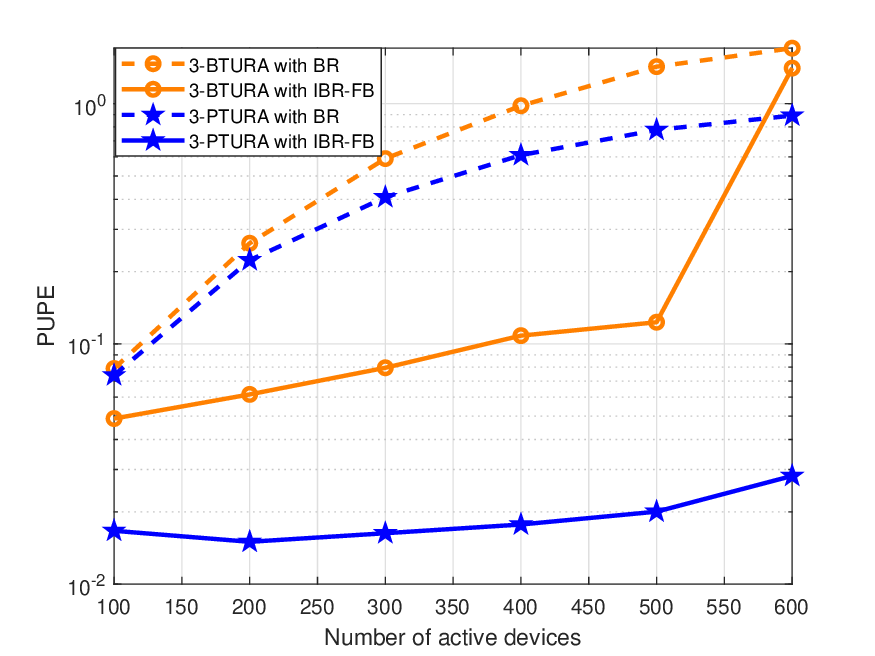}
	\caption{PUPE performance comparision between the 3-BTURA and the 3-PTURA, where $\frac{E_{\rm{b}}}{N_0}=-10$dB.}
	\label{fig_sim_polar_vs_bch}
\end{figure}

In this section, we clarify the advantages of the proposed coding scheme in the PTURA.
The benchmark for comparison is the BCH codes, with parameters prescribed by in \cite{p_tb_ura} (i.e., the active device message of length 96 is encoded to the BCH codeword of length 110.). 
At the transmitter, we substitute the CRC and Polar codes employed in the PTURA with the BCH code, thereby creating a modified scheme referred to as the BTURA. 
Concurrently, we propose an iterative Bayesian receiver with feedback (IBR-FB) for the BTURA, leveraging the inherent error detection capabilities of the BCH code.

We compare the performance of the 3-BTURA and the 3-PTURA in Fig. \ref{fig_sim_polar_vs_bch}, where we adopt the settings $[T_1,T_2,T_3]=[20,16,10]$ and $[B_1,B_2,B_3]=[48,38,24]$ for the 3-BTURA. To provide a comprehensive evaluation of the IBR-FB, we have also included in figure the performance of the Bayesian receiver without feedback (BR), which corresponds to the initial iteration of the IBR-FB. As observed from the results,
the PTURA demonstrates superior performance compared to the BTURA, particularly when the number of active devices is substantial. 
The performance gap between the PTURA and the BTURA stems from the fact that the PTURA takes use of the soft information  (i.e., the LLR $\mymat{r}_k$) fully.
In addition, the IBR-FB performs much better than the BR in both schemes. But we should recognize that the performance improvement of the IBR-FB comes at the cost of higher computational complexity.

\begin{figure}[!t]
	\centering
	\begin{subfigure}[b]{0.94\linewidth}
		\includegraphics[width=\linewidth]{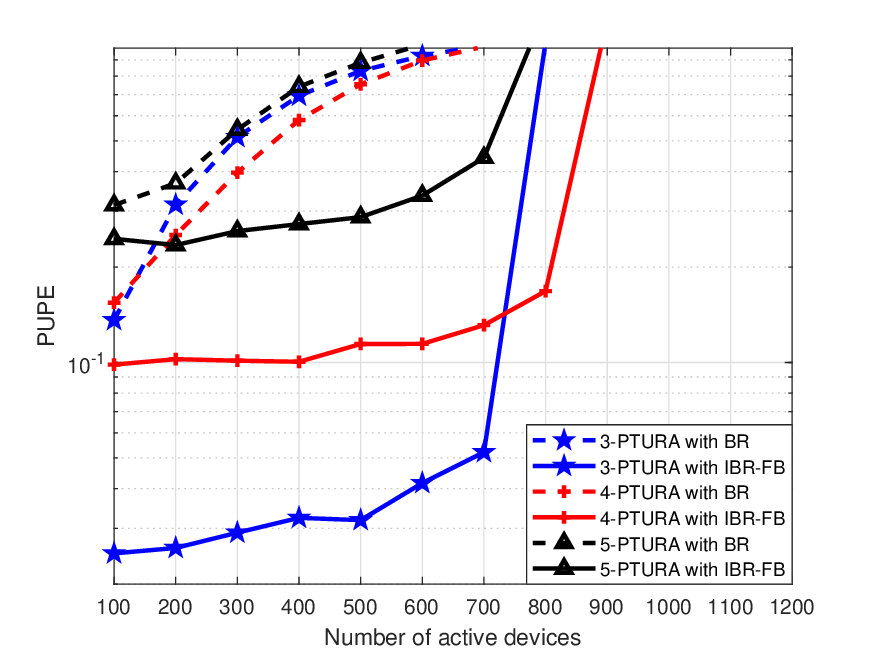}
		\caption{$\frac{E_{\rm{b}}}{N_0}=-10.5$dB.}
	\end{subfigure}
	\begin{subfigure}[b]{0.94\linewidth}
		\includegraphics[width=\linewidth]{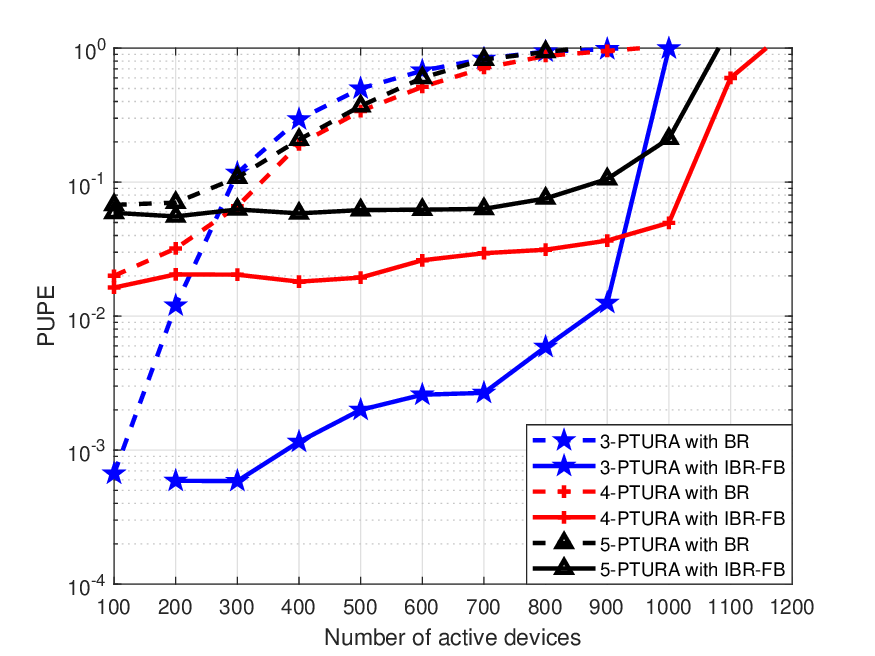}
		\caption{$\frac{E_{\rm{b}}}{N_0}=-7.5$dB.}
	\end{subfigure}
	\begin{subfigure}[b]{0.94\linewidth}
		\includegraphics[width=\linewidth]{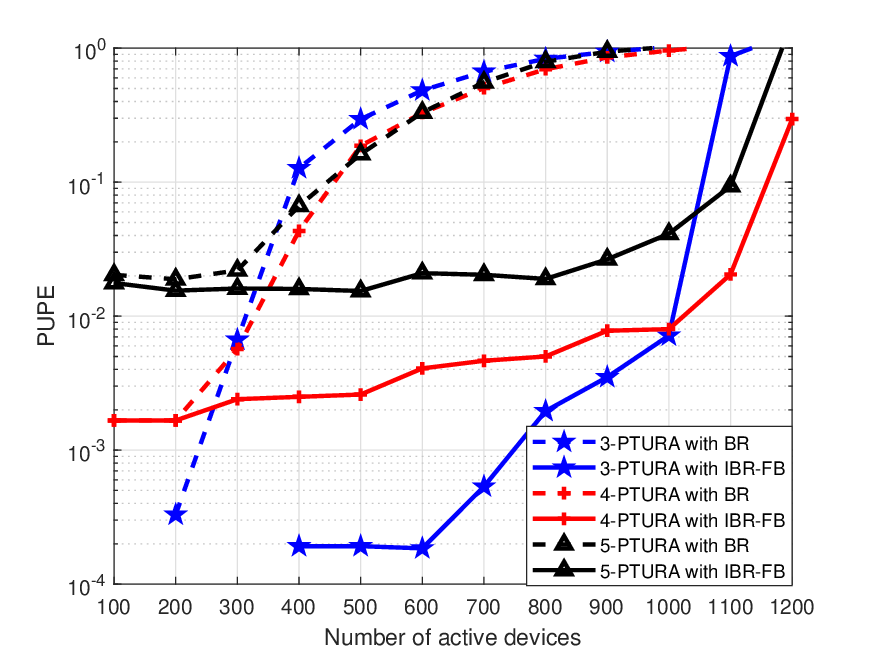}
		\caption{$\frac{E_{\rm{b}}}{N_0}=-4.5$dB.}
	\end{subfigure}
	\caption{PUPE performance of the PTURA.}
	\label{fig_sim_ds}
\end{figure}

\subsection{Evaluation of the PTURA Scheme}

In this section, we delve into the analysis of the system performance of the PTURA under varying parameter configurations. 
Fig. \ref{fig_sim_ds} compares the performance of the 3-PTURA, 4-PTURA, and 5-PTURA.
In scenarios with a smaller number of active devices, the performance of the 3-PTURA surpasses that of the 4-PTURA. However, when considering a relatively larger number of active devices, the 4-PTURA demonstrates superior performance compared to the 3-PTURA. 
The discrepancy between the TD performance and the decoding performance is what causes this phenomenon. 
Given a fixed block length $T$, the fewer of the number of segments $L$, the higher the freedom of the coding \cite{p_grass}, and thus the better the decoding performance. On the other hand, with the fixed block length $T$, the more number of segments, the greater the rank of the supported tensor that can be uniquely decomposed \cite{p_an_algorithm,p_onge}, and thus the larger of the supported number of active devices. 
In scenarios with a small number of active devices, the system performance is primarily determined by decoding performance, whereas in cases with a larger number of active devices, TD performance predominantly influences the system performance.
Nonetheless, the reduced coding freedom leads to the 5-PTURA underperforming in comparison to the 4-PTURA, irrespective of whether the number of active devices is small or large. This finding suggests that, when the number of segments is equal to or greater than 4, the PTURA cannot support more active devices by increasing the number of segments.
Compared with the PTURA with BR, it is evident that the PTURA with IBR-FB exhibits a substantial enhancement in performance. Moreover, the system performance of the PTURA with IBR-FB experiences a significantly slower decline as the number of active devices escalates.

\begin{figure}[!t]
	\centering 
	\includegraphics[width=0.47\textwidth]{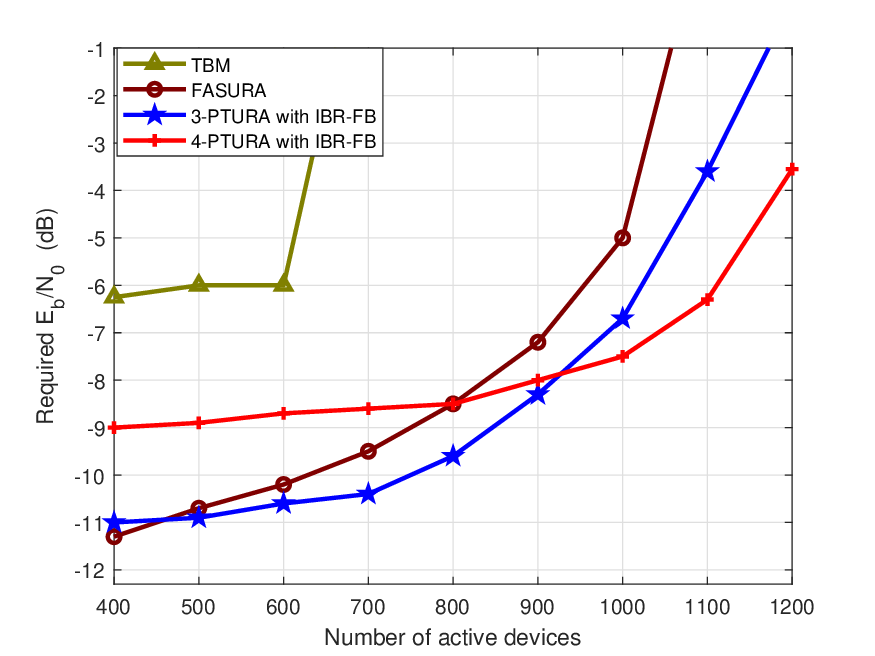}
	\caption{The required $\frac{E_{\rm{b}}}{N_0}$ when $\text{PUPE}=0.05$.}
	\label{fig_sim_edn}
\end{figure}

In Fig. \ref{fig_sim_edn}, we compare the required $\frac{E_{\rm{b}}}{N_0}$ of the PTURA with IBR-FB, the TBM proposed in \cite{p_tb_ura}, and the state-of-the-art URA scheme FASURA \cite{p_fas_ura},
when $\text{PUPE}=0.05$. 
The proposed PTURA demonstrates a significant performance gain compared to the TBM.
Furthermore, it is observed that the proposed 3-PTURA, with considerably less computational complexity (See Fig. \ref{fig_com}), outperforms FASURA when the number of active devices $K_{\rm{a}}$ exceeds 500. 
The performance gap between the two schemes widens as $K_{\rm{a}}$ increases. For instance, when $K_{\rm{a}}=500$, the gap between the two schemes amounts to 0.2 dB, and when $K_{\rm{a}}=1100$, the gap extends to more than 3.6 dB.
Indeed, the FASURA operates as an TURA scheme with $L=1$, which grants it greater coding freedom compared to the 3-PTURA. As a result, the FASURA can employ more powerful coding and modulation schemes (e.g., Polar coding with a lower code rate and random spreaing in the FASURA), and the FASURA performs better than the 3-PTURA when $K_{\rm{a}}\le 400$. However, this comes at the cost
of high computational complexity.
Furthermore, when comparing the 3-PTURA and 4-PTURA, a phenomena consistent with the observations in Fig. \ref{fig_sim_ds} becomes evident.
If $K_{\rm{a}}\ge 900$, then the 4-PTURA performs better than the 3-PTURA; otherwise, the 3-PTURA is better than the 4-PTURA. 

\section{Conclusion}

\label{sec_conclusion}
In this paper, we investigated the PTURA scheme. The IBR-FB was proposed for the PTURA, relying on the error detection capabilities of the CRC codes.
	Specifically, under the variational Bayesian learning framework, we proposed the GM-BTD algorithm for exporting the required soft information without the knowledge of the number of active devices. Subsequent to this, we developed the soft Grassmannian demodulator to compute the required LLR.
	Numerous simulations demonstrated that the proposed PTURA scheme outperforms the existing state-of-the-art unsourced random access scheme.

\appendices

\normalem

\bibliographystyle{IEEEtran}
\bibliography{mybibfile}

\end{document}